\documentclass[12pt,aip,jcp,reprint]{revtex4-1}

\usepackage[utf8]{inputenc}
\usepackage[UKenglish]{babel}
\usepackage{fontenc}            	% tells LaTeX how to output the text

\usepackage{amsmath}            	% amsmath package
\usepackage{amsfonts}           	% amsmath fonts
\usepackage{amssymb}            	% amsmath symbole
\usepackage{mathrsfs}           	% other mathematical symbols
\everymath{\displaystyle}

\usepackage{graphicx}          		% manage pictures
\usepackage{caption}				% figure captions
\usepackage{subcaption}				% figure subcaptions
\usepackage{floatrow}           	% place a caption beside a float
\usepackage{multirow}           	% spanning rows in tables
\usepackage{float}              	% figures with borders
\floatstyle{plaintop}           	% captions above the float (or plain, boxed, ruled)
\restylefloat{table}            	% apply the style to tables
\usepackage[usenames]{color}    	% used for font color
\usepackage[svgnames]{xcolor}		% colors by their 'svgnames'
\usepackage[pdftex,%
			colorlinks,%
			bookmarksopen,%
			bookmarksnumbered,%
			citecolor=black,%
			linkcolor=black,%
			urlcolor=black]{hyperref}   		% hyper reference
\usepackage{ulem}

\newcommand{\dd}[1]{\,\mathrm{d}#1\,}      	% integral d symbol
\newcommand{\bb}[1]{\left(#1\right)}        % normal brackets
\newcommand{\bba}[1]{\left<#1\right>}       % angle brackets
\newcommand{\bbm}[1]{\left|#1\right|}       % modulus brackets
\newcommand{\bbs}[1]{\left[#1\right]}       % square brackets
\newcommand{\lint}[2]{\int\limits_{#1}^{#2}}        	% single int limits
\newcommand{\lsum}[2]{\sum\limits_{#1}^{#2}\,}   		% single int limits

\newcommand{\etal}{\textit{et al}.}                 	% et al citation
\newcommand{\fig}[1]{Fig.~\ref{#1}}               		% figure reference
\newcommand{\eqn}[1]{Eq.~(\ref{#1})}                	% equation reference
\newcommand{\Eqn}[1]{Equation~(\ref{#1})}         		% equation reference
\begin{document}


%\author{}
%\altaffiliation[Present address: ]{}
%\email[]{}
%\thanks{}
%\affiliation{}

\title{The pressure tensor across a liquid-vapour interface}

\author{Carlos Braga}
\email[]{ccorreia@imperial.ac.uk}
\affiliation{Department of Chemical Engineering, Imperial College London, SW7 2AZ UK}

\author{Edward Smith}
\email[]{edward.smith05@imperial.ac.uk}
\affiliation{Department of Chemical Engineering, Imperial College London, SW7 2AZ UK}

\author{Andreas Nold}
\email[]{andreas.nold09@imperial.ac.uk}
\affiliation{Department of Chemical Engineering, Imperial College London, SW7 2AZ UK}

\author{David N. Sibley}
\email[]{d.n.sibley@lboro.ac.uk}
\affiliation{Department of Mathematical Sciences, Loughborough University, Leicestershire LE11 3TU UK.}

\author{Serafim Kalliadasis}
\email[]{s.kalliadasis@imperial.ac.uk}
\affiliation{Department of Chemical Engineering, Imperial College London, SW7 2AZ UK}

\date{\today}

\begin{abstract}
Inhomogeneous fluids exhibit physical properties that are neither uniform nor
isotropic. The pressure tensor is a case in point, key to the mechanical
description of the interfacial region. Kirkwood and Buff, and later Irving
and Kirkwood, obtained a formal treatment based on the analysis of the
pressure across a planar surface ~[J.G. Kirkwood and F.P. Buff,
J.~Chem.~Phys., 17(3), (1949), J.H. Irving and J.G. Kirkwood, J.~Chem.~Phys.
18, 817 (1950)]. We propose a generalisation of Irving and Kirkwood's
argument to fluctuating, non-planar surfaces and obtain an expression for the
pressure tensor that is not smeared by thermal fluctuations at the molecular
scale and corresponding capillary waves ~[F.P. Buff, R.A. Lovett, and F.H.
Stillinger, Jr., Phys.~Rev.~Lett. 15, 621 (1965)]. We observe the emergence
of surface tension, defined as an excess tangential stress, acting exactly
across the dividing surface at the sharpest molecular resolution. The new
statistical mechanical expressions extend current treatments to fluctuating
inhomogeneous systems far from equilibrium.
\end{abstract}
         	% use showkeys class option if keyword
\maketitle          	%\maketitle must follow title, authors, abstract and \pacs

%%%%%%%%%%%%%%%%%%%%%%%%
% Body of the paper
% Long equations
% If in two-column mode, this environment will change to single-column format so that long equations can be displayed.
%\begin{widetext}
%$$\mbox{put long equation here}$$
%\end{widetext}
%
% Figures
% Figures should be put into the text as floats.
% Use the graphics or graphicx packages (distributed with LaTeX2e).
% Put the label that you will use with \ref{} command in the braces of the \label{} command.
%
% \begin{figure}
% \end{figure}
%
% Tables
% \begin{table}
% \end{tabular}
% \end{table}
%
% Acknowledgements
%\begin{acknowledgments}
% Put your acknowledgments here.
%\end{acknowledgments}

%%%%%%%%%%%%%%%%%%%%%%%%
\section{Introduction}
\label{sec:intro}
% Background
Inhomogeneous systems are characterised by an interface, across which
physical quantities vary sharply, over a length scale of a few molecular
diameters~\citep{Rowlinson2002aa}. Thermally induced fluctuations of the
interface in the form of capillary waves smooth out the transition of
properties across the interface. Although capillary wave behaviour has been
observed experimentally~\citep{Sanyal1991aa,Tidswell1991aa}, analysis of the
interfacial structure at the sharpest molecular resolution is difficult as a
result of these fluctuations. Capillary wave theory (CWT), introduced by Buff
\etal~\citep{Buff1965ab,Evans1979aa}, accounts for these by assuming the
existence of an intrinsic surface $\textstyle\xi\bb{\mathbf{x}}$, where
$\mathbf{x}$ is a vector in the coordinate system parallel to the surface.
This function acts as an instantaneous boundary between fluid phases, against
which statistical averages are computed, e.g. the intrinsic density profile,
\begin{equation}
\label{eqn:intro1}
\tilde{\rho}\bb{z+\xi} = \frac{1}{A}\lsum{i=1}{N}\bba{\delta\bb{z+\xi\bb{\mathbf{x}_{i}}-z_{i}}}
\end{equation}
\noindent where $\bb{\mathbf{x}_{i},z_{i}} = \bb{x_{i}, y_{i}, z_{i}}$ are
the atom positions and $A=L_{x}L_{y}$ is the transverse area. In its original
form~\citep{Buff1965ab}, CWT sees the interface at the largest wavelength as
a step function, whose local position is described by a Gaussian distribution
$P\bb{\xi}\propto\exp(-(\xi-\bba{\xi})^{2}/2w^{2})$. The square width $w^2$
characterises the correlation length of the surface fluctuations and is known
to have the taxing property of, in absence of an external field, diverging in
the thermodynamic limit~\citep{Percus1986aa,Fernandez2013aa}. The connection
with the singlet density $\rho\bb{z}$ is often accomplished by ignoring the
coupling between capillary wave fluctuations and interfacial structure,
resulting in a congruent fluctuation of the intrinsic profile. The mean
density profile is then obtained from the convolution over the capillary
waves, $\textstyle\rho\bb{z} =
\int\dd{\xi}\tilde{\rho}\bb{z-\xi}\,P\bb{\xi}$~\citep{Muller2005aa}.

Among the different results reported previously, the interfacial
Hamiltonian approach
(e.g.~Refs~\onlinecite{Dietrich1999,Parry2000,Dietrich2015,Peter2016})
gives useful information on the wavenumber dependence of surface tension
but relies on various phenomenological parameters which are not known before hand.
Further recently, there has been a renewed interest in the analysis of
the pressure tensor in inhomogeneous systems, notably by
Sega~\etal~in~\citep{Sega2015aa} for a Lennard-Jones (LJ) fluid and
in~\citep{Sega2016ab} for molecular liquids. In these works the authors
performed a deconvolution of the pressure tensor from the fluctuations in the
interface position, thereby obtaining its intrinsic counterpart. The
intrinsic profile of single particle properties is trivially obtained by
appropriate weighting of the sum in~\eqn{eqn:intro1},
$\textstyle\tilde{\rho}_{w}(z)=\sum_{i}\,\bba{w_{i}\delta(z +
\xi\bb{\mathbf{x}_{i}} - z_{i})}/A$, where $w_{i}$ is an arbitrary atomic
weight. In contrast,  the potential component of the pressure,
$\mathbf{P}^{c}(\mathbf{r})$, is not easily defined due to the intermolecular
interactions~\citep{Todd1995aa}. For a pairwise potential,
$\textstyle\mathbf{P}^{c}(\mathbf{r})=1/2\sum_{i\neq
j}\,\bba{\mathbf{F}_{ij}\int_{C_{ij}}\dd{\mathbf{l}}\delta(\mathbf{l}-\mathbf{r})}$
where $\mathbf{F}_{ij}$ is the force between the pair of atoms, and $C_{ij}$
is the path of the line integral connecting atom $i$ and $j$.

There is no unique way of defining $C_{ij}$, and so there is also no unique
way of determining which atom pairs contribute to the force across a given
surface element~\citep{Baus1990aa}. Irving-Kirkwood (IK)
convention~\citep{Irving1950aa} stipulates that the force between two atoms
is said to be \textit{across} if the vector connecting the centres of mass of
the two atoms intersects the surface element. The alternative Harasima (H)
contour~\citep{Harasima1958aa} breaks the line connecting a pair of atoms
into two mutually orthogonal paths. For planar interfaces, the H
representation is seen as particularly convenient, because by construction,
the transverse contribution of each pair of particles becomes path
independent~\citep{Sega2015aa}; on the other hand, there are always two
possible H contours connecting a pair of particles, first transverse and then
normal $\uparrow\circ\rightarrow$, and vice versa,
$\rightarrow\circ\uparrow$. In the presence of an interface, this distinction
is important, because it defines the position where the contribution to total
pressure is accounted for. This is usually mitigated by equally partitioning
the contribution at each particle's location but it does not resolve the
fundamental ambiguity.

% Motivation
Furthermore, it is also recognised that the normal
component of this contour is known to not be physically meaningful~\citep{Sega2015aa,Schofield1982aa}.
This additional conceptual difficulty is usually reconciled by arguing that mechanical stability,
$\textstyle\mathbf{\nabla}\cdot\mathbf{P}=0$,
requires uniformity of the normal pressure across the system.
As it stands, this assumption can only hold under equilibrium conditions.
Gradients in mass, momentum and energy will generate thermodynamic fluxes
that change the liquid structure, vis-a-vis the interfacial properties~\citep{Todd2007aa,Evans2008aa}.
Physical modelling of coupling between these is key to understanding the
behaviour of fluids under non equilibrium conditions.
Examples include the suppression of thermal fluctuations by shear flow~\citep{Derks2006aa,Thiebaud2010aa},
Faraday instability at the liquid-vapour interface~\citep{Francois2017aa} and
Marangoni flow~\citep{Miniewicz2016aa}.

The motivation of this work is thus to extend the IK definition of the
pressure tensor as a force acting across a non-planar surface, fluctuating as
a result of the thermal motions of the fluid. Starting from the continuity
equations of hydrodynamics, expressed relative to a fluctuating reference
frame, new expressions are derived which are valid for any inhomogeneous
fluid far from equilibrium. Unlike H or IK line integrals, the new contour of
integration is a dynamic quantity, reflecting the instantaneous local
curvature of the interface. Using the control volume
approach~\cite{Smith2012aa} to the IK equations, we compute the pressure
profile of a well known liquid-vapour LJ fluid. We find the emergence of the
surface tension acting exactly across the intrinsic surface, free from the
smearing effects of the surface separating the liquid and vapour phases.

% Organisational
%The article is organised as follows, first the theoretical background will be developed,
%by introducing CWT and the IK pressure tensor.
%These will then be combined to demonstrate the derivation of the intrinsic density from first principles.
%The same process will then be applied to derive the corresponding intrinsic stress tensor.
%The methodology section outlines the simulation details.
%The results section presents and discusses the new intrinsic density and pressure profiles of a Lennard-Jones system.
%Finally, the conclusions of this work are presented.

%%%%%%%%%%%%%%%%%%%%%%%%
\section{Intrinsic Pressure Tensor}
\label{sec:theory}
In its original form, IK microscopic expressions for thermodynamic fluxes are written in terms of ensemble averages.
Thermodynamic fluxes are, however, based on conservation laws, valid instantaneously over every member in the
statistical ensemble. The Dirac delta functions become a convenient operator whose physical meaning only
becomes apparent when averaging over the volume~\citep{Evans2008aa}.
We therefore drop the ensemble averaging of the density operators in our analysis
following the procedure of Todd~\etal~\citep{Todd2007aa,Todd1995aa}.
The mass density $\rho\bb{\mathbf{r},t}$ and momentum
density  $\mathbf{J}\bb{\mathbf{r},t}$ are defined by,
\begin{align}
\label{eqn:dens1}
\rho\bb{\mathbf{r},t}
&= \lsum{i=1}{N}m_{i}\,\delta\bb{\mathbf{r}-\mathbf{r}_{i}\bb{t}}\;,\\
\label{eqn:dens2}
\mathbf{J}\bb{\mathbf{r},t}
&= \lsum{i=1}{N}m_{i}\dot{\mathbf{r}}_{i}\,\delta\bb{\mathbf{r}-\mathbf{r}_{i}\bb{t}}\;,
\end{align}
\noindent where $m_{i}$, $\mathbf{r}_{i}$ and $\dot{\mathbf{r}}_{i}$ are the
particles' mass, position and total velocity respectively. The hydrodynamic
formulation~\cite{Smith2012aa} of the these allows the densities to be
expressed in control volume form rendering them more analytically tractable.
Given a region of interest, the integral of the local density
\begin{equation}
\label{eqn:dens3}
\lint{V}{}\dd{\mathbf{r}}\rho\bb{\mathbf{r},t} =
\lint{V}{}\dd{\mathbf{r}}\lsum{i=1}{N}m_{i}\,\delta\bb{\mathbf{r}-\mathbf{r}_{i}}\;,
\end{equation}
\noindent is the total mass inside the control volume. The region interval is
arbitrary as long as it is compact. Assuming that one can obtain the
intrinsic surface $\textstyle\xi\bb{\mathbf{x}}$ over the $xy$-plane for a
given configuration, the volume can be foliated into regular elements
\begin{equation}
\label{eqn:dens4}
\lint{V}{} \dd{\mathbf{r}} =
\lint{x^{-}}{x^{+}}\dd{x}
\lint{y^{-}}{y^{+}}\dd{y}
\lint{z^{-}+\xi\bb{x,y}}{z^{+}+\xi\bb{x,y}}\dd{z} =
\Delta x\,\Delta y\,\Delta z\;,
\end{equation}
\noindent where the limits of integration are defined as $\Delta \mathbf{r} =
\mathbf{r}^{+} - \mathbf{r}^{-}$. Here,  we are interested in the
distribution of the pressure as a function of the distance to the interface.
This can be achieved by a change of coordinate system and, in accordance with
CWT assumptions, we adopt the view that surface fluctuations are independent
of interfacial structure. The arbitrary surface
$\textstyle\xi\bb{\mathbf{x}}$ is therefore taken to be independent of time,
thereby avoiding the appearance of time dependent contributions to the
pressure due to the dynamic motion of the surface. Integration of the
operators $\delta\bb{\mathbf{r} - \mathbf{r}_i}$ over the control volume has
the effect of isolating the particles inside the  region of
interest~\cite{Smith2012aa}, and \eqn{eqn:dens3} expands to
\begin{widetext}
\begin{equation}
\begin{aligned}
\label{eqn:dens5}
\lint{V}{}\dd{\mathbf{r}}\rho\bb{\mathbf{r},t}
&=
\lint{x^{-}}{x^{+}}\dd{x}
\lint{y^{-}}{y^{+}}\dd{y}
\lint{z^{-}+\xi\bb{x,y}}{z^{+}+\xi\bb{x,y}}\dd{z}
\lsum{i=1}{N}m_{i}\,\delta\bb{\mathbf{r}-\mathbf{r}_{i}}\\
%&=
%\lint{x^{-}}{x^{+}}\dd{x}
%\lint{y^{-}}{y^{+}}\dd{y}
%\lsum{i=1}{N}m_{i}\,
%\delta\bb{x-x_{i}}\,\delta\bb{y-y_{i}} \vartheta_{z_{i}}\bbs{z^{-} + \xi\bb{x,y}, z^{+} + \xi\bb{x,y}}\\
&=
\lsum{i=1}{N}m_{i}\,
\vartheta_{x_{i}}\bbs{x^{-}, x^{+}}
\vartheta_{y_{i}}\bbs{y^{-}, y^{+}}
\vartheta_{z_{i}}\bbs{z^{-} + \xi\bb{x_{i},y_{i}}, z^{+} + \xi\bb{x_{i},y_{i}}}\;,
\end{aligned}
\end{equation}
\end{widetext}
\noindent where the $x$ and $y$ integrals, in the last line, use the sifting property
$\textstyle\int_{-\infty}^\infty \dd{\alpha} f(\alpha)\delta(\alpha-\alpha_i) = f(\alpha_i) $;
the boxcar function $\vartheta$ results from the integral of the Dirac delta between finite limits,
$\vartheta_{\alpha_{i}}[\alpha^{-},\alpha^{+}] \equiv \theta\bb{\alpha^{+} - \alpha_{i}}-\theta\bb{\alpha^{-} - \alpha_{i}}$,
$\alpha=x,y,z$ and $\theta\bb{\alpha}$ is the Heaviside step function.
Assuming periodic boundary conditions in the transverse directions,
$\vartheta_{x_{i}} \!\bbs{0, L_{x}} = 1 \; \forall \; x_i \in (0, L_x) $ and
$\vartheta_{y_{i}} \!\bbs{0, L_{y}} = 1 \; \forall \; y_i \in (0, L_y)$,
\begin{equation}
\label{eqn:dens6}
\lint{V}{}\dd{\mathbf{r}}\rho\bb{\mathbf{r},t} =
\lsum{i=1}{N}m_{i} \vartheta_{z_{i}}\bbs{z^{-} + \xi\bb{\mathbf{x}_{i}}, z^{+} + \xi\bb{\mathbf{x}_{i}}},
\end{equation}
\noindent where $\mathbf{x}_{i}=\bb{x_{i}, y_{i}}$.
The mean value theorem applied on the left hand side of \eqn{eqn:dens6}
gives a measure of the volume average density
$\textstyle \tilde{\rho} (z\,+\,\xi) \approx \int_V\dd{\mathbf{r}}\rho\bb{\mathbf{r},t} / \Delta V$,
$z \in (z^{-}, z^{+})$ with $\Delta V = A \,\Delta z$.
Using the definition of the Dirac delta function as the zero volume limit $\Delta z \to 0$,
$\textstyle\delta\bb{z} \equiv \lim_{\Delta z\to 0}\bb{\theta\bb{z + \Delta z/2} - \theta\bb{z-\Delta z/2}}/\Delta z$,
the density at given position relative to the surface, $z+\xi\bb{\mathbf{x}}$ is given by
\begin{equation}
\label{eqn:dens7}
\tilde{\rho}\bb{z+\xi} =
\frac{1}{A}\lsum{i=1}{N}m_{i}\,\delta\bb{z + \xi\bb{\mathbf{x}_{i}} - z_{i}}\;.
\end{equation}
The above equation defines the instantaneous density relative to a given surface.
Assuming that one can somehow compute $\textstyle\xi\bb{\mathbf{x}}$, the average over the ensemble
of configurations and corresponding surfaces is the intrinsic density profile, $\bba{\tilde{\rho}\bb{z+\xi}}$~\citep{Buff1965ab}.
The above equation deserves a particular emphasis,
because its  derivation is based on a microscopic expression of a hydrodynamic quantity,
without any assumption regarding the distribution function $\textstyle f\bb{\mathbf{r}^{N},\mathbf{p}^{N}, t}$.
It can therefore form the basis for a statistical mechanical derivation of the pressure tensor
in a Lagrangian frame of reference relative to the instantaneous surface.
Starting from the definition of momentum, \eqn{eqn:dens2}, the same process can be followed to
obtain an instantaneous momentum density relative to the surface,
\begin{equation}
\label{eqn:dens8}
\tilde{\mathbf{J}}\bb{z+\xi} =
\frac{1}{A}\lsum{i=1}{N}m_{i} \dot{\mathbf{r}}_{i} \,\delta\bb{z + \xi\bb{\mathbf{x}_{i}} - z_{i}}\;.
\end{equation}
The time derivative of \eqn{eqn:dens2} at a given point results in the momentum conservation
equation where, for a system of pairwise interacting particles, the microscopic representation
of the IK pressure tensor~\citep{Evans2008aa} is given by
\begin{equation}
\label{eqn:pres1}
\begin{split}
\mathbf{P}\bb{\mathbf{r},t}
={} & \mathbf{P}^{k}\bb{\mathbf{r},t} + \mathbf{P}^{c}\bb{\mathbf{r},t}\\
=& \lsum{i=1}{N}m_{i}\mathbf{v}_{i}\mathbf{v}_{i}\,\delta\bb{\mathbf{r}-\mathbf{r}_{i}}\\
 &+\frac{1}{2}\lsum{\substack{i,j=1\\i\neq j}}{N}\mathbf{r}_{ij}\mathbf{F}_{ij}\,\lint{0}{1}\dd{\lambda}\delta\bb{\mathbf{r}-\mathbf{r}_{\lambda}}\;,
\end{split}
\end{equation}
\noindent where $\mathbf{P}^{k}\bb{\mathbf{r},t}$ is the kinetic contribution to the pressure,
obtained by decomposition of the total velocity $\dot{\mathbf{r}}_{i}$ into a thermal component
$\mathbf{v}_{i}$ and a streaming part $\mathbf{u}\bb{\mathbf{r}_{i},t}$, taken to be zero in the current work,
$\dot{\mathbf{r}}_{i}=\mathbf{v}_{i}+\mathbf{u}\bb{\mathbf{r}_{i},t}$.
Akin to density (cf. \eqn{eqn:dens1}) and momentum (cf. \eqn{eqn:dens2}),  $\mathbf{P}^{k}\bb{\mathbf{r},t}$ is localised at the atom positions by $\delta\bb{\mathbf{r}-\mathbf{r}_{i}}$.
The configurational part of the pressure tensor, $\mathbf{P}^{c}\bb{\mathbf{r},t}$,
is obtained by integrating along the line connecting a pair of atoms, where each point
$\mathbf{r}_{\lambda} = \mathbf{r}_{i} - \lambda\mathbf{r}_{ij}$,
is sampled by $\delta\bb{\mathbf{r}-\mathbf{r}_{\lambda}}$.
Starting from the IK form of the pressure in \eqn{eqn:pres1} and following the same procedure used to obtain intrinsic density above,
we aim to obtain the IK pressure tensor mapped onto a coordinate system defined by the instantaneous intrinsic surface.
The integral of the pressure tensor over a control volume bounded by a surface $\textstyle\xi\bb{\mathbf{x}}$ is given by
\begin{widetext}
\begin{equation}
\label{eqn:pres2}
\begin{split}
\lint{V}{}\dd{\mathbf{r}}\mathbf{P}\bb{\mathbf{r},t}
=&
\lint{x^{-}}{x^{+}}\dd{x}
\lint{y^{-}}{y^{+}}\dd{y}
\lint{z^{-}+\xi\bb{x,y}}{z^{+}+\xi\bb{x,y}}\dd{z}
\lsum{i=1}{N}m_{i}\mathbf{v}_{i}\mathbf{v}_{i}\,\delta\bb{\mathbf{r}-\mathbf{r}_{i}}\\
&+\frac{1}{2}
\lint{x^{-}}{x^{+}}\dd{x}
\lint{y^{-}}{y^{+}}\dd{y}
\lint{z^{-}+\xi\bb{x,y}}{z^{+}+\xi\bb{x,y}}\dd{z}
\lsum{\substack{i,j=1\\i\neq j}}{N}\mathbf{r}_{ij}\mathbf{F}_{ij}\,\lint{0}{1}\dd{\lambda}\delta\bb{\mathbf{r}-\mathbf{r}_{\lambda}}\\
=&
\lsum{i=1}{N}m_{i}\mathbf{v}_{i}\mathbf{v}_{i}\,
\vartheta_{x_{i}}\bbs{x^{-}, x^{+}}
\vartheta_{y_{i}}\bbs{y^{-}, y^{+}}
\vartheta_{z_{i}}\bbs{z^{-} + \xi\bb{x_{i},y_{i}}, z^{+} + \xi\bb{x_{i},y_{i}}}\\
&+\frac{1}{2}
\lsum{\substack{i,j=1\\i\neq j}}{N}\mathbf{r}_{ij}\mathbf{F}_{ij}\,\lint{0}{1}\dd{\lambda}
\vartheta_{x_{\lambda}}\bbs{x^{-}, x^{+}}
\vartheta_{y_{\lambda}}\bbs{y^{-}, y^{+}}
\vartheta_{z_{\lambda}}\bbs{z^{-} + \xi\bb{x_{\lambda},y_{\lambda}}, z^{+} + \xi\bb{x_{\lambda},y_{\lambda}}}\;,
\end{split}
\end{equation}
\end{widetext}
\noindent where $x_{\lambda}$, $y_{\lambda}$ and $z_{\lambda}$ are the components
of a point in the vector connecting particles $i$ and $j$.
The integral of $\lambda$ from $0$ to $1$ moves along the line and partitions the
energy between the different volume elements across which the contour line passes through.
This seemingly complicated looking function has the clear geometric meaning that the
path between each pair of particles is not a straight line but a curve
-- unique to each pair of interacting particles -- whose coordinates
are dictated by the shape of the interface at the transversal coordinates $x_{\lambda}$ and $y_{\lambda}$.
Approximating the energy in \eqn{eqn:pres2} by its volume average,
$\textstyle\tilde{\mathbf{P}}\bb{z\,+\,\xi}\approx\int_V\dd{\mathbf{r}}\mathbf{P}\bb{\mathbf{r},t}/\Delta V$,
over the $xy$-plane, and using the Heaviside definition of the Dirac delta function when $\Delta z \to 0$,
the instantaneous pressure relative to the surface is given by

\begin{equation}
\label{eqn:pres3}
\begin{split}
\tilde{\mathbf{P}}\bb{z+\xi}
=& \tilde{\mathbf{P}}^{k}\bb{z+\xi} + \tilde{\mathbf{P}}^{c}\bb{z+\xi}\\
=& \frac{1}{A}\lsum{i=1}{N}m_{i}\mathbf{v}_{i}\mathbf{v}_{i}\,\delta\bb{z + \xi\bb{\mathbf{x}_{i}} - z_{i}}\\
&+\frac{1}{2A} \lsum{\substack{i,j=1\\ i\neq j}}{N}\mathbf{r}_{ij}\mathbf{F}_{ij}\,\lint{0}{1}\dd{\lambda}
\delta\bb{z + \xi\bb{\mathbf{x}_{\lambda}} - z_{\lambda}}\;,
\end{split}
\end{equation}
\noindent where $\mathbf{x}_{\lambda}=\bb{x_{\lambda}, y_{\lambda}}$ are the coordinates of the line
between a pair of particles transversal to the interface.
In the particular case where the intrinsic surface is planar, $\xi\bb{\mathbf{x}} = c$, which can be absorbed into
the coordinate position $z$, and the above equation gives the IK expression for the pressure tensor, \eqn{eqn:pres1},
averaged over the $xy$-plane,
\begin{equation}
\label{eqn:pres4}
\begin{split}
\tilde{\mathbf{P}}\bb{z}
=& \tilde{\mathbf{P}}^{k}\bb{z} + \tilde{\mathbf{P}}^{c}\bb{z}\\
=& \frac{1}{A}\lsum{i=1}{N}m_{i}\mathbf{v}_{i}\mathbf{v}_{i}\,\delta\bb{z - z_{i}}\\
&+\frac{1}{2A} \lsum{\substack{i,j=1\\ i\neq j}}{N}\mathbf{r}_{ij}\mathbf{F}_{ij}\,\lint{0}{1}\dd{\lambda}
\delta\bb{z - z_{\lambda}}\;.
\end{split}
\end{equation}
\Eqn{eqn:pres3} is the main result of the current work.
Unlike IK and H conventions, the contour line for each pair of particles is a unique path that reflects the
local curvature of the instantaneous interface, setting the current analysis apart from previous studies~\citep{Hardy1982aa,Lutsko1988aa,Sega2016ab,Sega2015aa}.
Its average over an ensemble of configurations gives a precise definition of the intrinsic pressure tensor,
$\tilde{\mathbf{P}}\bb{z+\xi}= \bba{\mathbf{P}\bb{z+\xi}}$, free from the smearing effect of surface fluctuations.
Because its foundations are hydrodynamic, it is not bound to any equilibrium distribution, thereby paving the way
to the study of complex fluids under non equilibrium conditions.

%%%%%%%%%%%%%%%%%%%%%%%%
\section{Simulation of Liquid-Vapour Interface}
\label{sec:methods}
\subsection{Intrinsic Pressure and Surface Tension}
We apply \eqn{eqn:pres3} to the analysis of a liquid-vapour interface of a
simple LJ fluid. Surface tension is the defining physical property
characterizing the interfacial structure. From a mechanical standpoint, it
arises from the anisotropy of the pressure tensor, reflecting the distinct
way molecular interactions vary across the system,
\begin{equation}
\label{eqn:tension1}
\gamma = \lim_{l\to\infty}\lint{-l}{l}\dd{z}\bb{P_{N}\bb{z} - P_{T}\bb{z}}\;,
\end{equation}
\noindent where $P_{N}$, and $P_{T}$ are the components of the pressure tensor normal and transverse to the surface (assuming cylindrical symmetry).
The link between these two approaches is given by the principle of virtual work, summarised here for completeness.
Given a fluid element $\mathrm{d}V=L_{x}\,L_{y}\mathrm{d}z=A\mathrm{d}z$ and a deformation tensor given by
$\epsilon_{xx}=\epsilon_{yy}=\bb{1+\epsilon}$,
$\epsilon_{zz}=1/(\epsilon_{xx}\,\epsilon_{yy})$ and
$\epsilon_{\alpha\beta}=0\,,\alpha\neq\beta$, then to first order in the Cauchy strain $\epsilon$, the change in area is
$\delta A=2\epsilon A$ and the work done by the system is
\begin{equation}
\label{eqn:tension2}
\delta W = \epsilon L_{x}\,L_{y}\dd{z}\bb{P_{x}\bb{z} + P_{y}\bb{z}  - 2P_{z}\bb{z}}\;.
\end{equation}
By virtue of its definition, the transformation preserves the volume, so the free energy cost per unit area is
$\delta F/\delta A=\mathrm{d}z\,(P_{N}(z)- P_{T}(z))$, where $P_{N}=P_{z}$ and $P_{T}=(P_{x}+P_{y})/2$.
The analysis merits some justification: (i) the approximation is only valid for infinitesimal deformations $\epsilon\ll 1$,
hinting at the fact surface tension is an equilibrium property and (ii) any excess pressure in the transversal direction will
cost energy wherever its position across the system.
What is perhaps unnoticed from this argument is that gradients in density may break the symmetry of the
pressure components and are bound to create tension in the direction transversal to the interface.
Although this has been observed previously~\citep{Sega2015aa},
one aim of the current study is to quantitatively analyse in detail this behaviour across the interfacial region.
In accordance with CWT, taking the interface at the longest wavelength, we can use \eqn{eqn:pres3} to decouple
the capillary wave fluctuations and use it to analyse the molecular origins of the surface tension at the interface,
\begin{equation}
\label{eqn:tension3}
\tilde{\gamma} = \lim_{l\to\infty}\lint{-l}{l}\dd{z}\bb{\tilde{P}_{N}\bb{z} - \tilde{P}_{T}\bb{z}}\;.
\end{equation}

\subsection{Molecular Dynamics Simulations}
Molecular dynamics (MD) simulations of a liquid-vapour interface are performed with a system of $N=6000$ particles inside
a simulation cell of dimensions $L_{x}=L_{y}=14.0$ and $L_{z}=80.0$, where unless otherwise indicated, reduced units are used throughout this work.
The system size was chosen to be larger than the minimum value of $N=1000$ particles as illustrated by
Trokhymchuk~\etal~\citep{Trokhymchuk1999aa}.
The shorter dimension of the simulation box was chosen to be larger than the  $L_{x}\approx 10$ to avoid spurious correlations
in the interface fluctuations~\citep{Bresme2008aa} and longer dimension $L_{z}>5L_{x}$ was chosen
to avoid coupling between the two interfaces in the system.
Atoms interact through a spherical truncated and shifted LJ potential with a cutoff radius of $r_{cut}=2.5$,
\begin{equation}
\label{eqn:meth1}
\phi\bb{r} = 4\epsilon\,\left[\left(\frac{\sigma}{r}\right)^{12}-\left(\frac{\sigma}{r}\right)^{6}\right]-\phi\bb{r_{cut}},\quad r<r_{cut}\;,
\end{equation}
\noindent where $r$ is the distance between a pair of particles and the corresponding energy and length
scales are defined by $\epsilon=1$ and $\sigma=1$, respectively.

Two temperatures are considered, $T=0.6$ and $T=0.7$. These were chosen to
emphasise the effects of a dense fluid with a sharp liquid-vapour interface.
Although different values of the triple point temperature of a LJ fluid have
been reported (e.g., Mastny and de Pablo give a value of
$T=0.694$~\citep{Mastny2007aa} while Errington \etal ~\citep{Errington2003aa}
report a value of $T=0.560$ using the same cutoff as the current work), it
has been shown previously~\citep{Braga2016aa} that the current pair radial
distribution functions of the studied systems do not exhibit evidence of
solid-phase structures. The equations of motion are integrated using the
LAMMPS software package~\citep{Plimpton1995aa}. The temperature is controlled
with a Nos\'{e}-Hoover~\citep{Hoover1985aa,Nose1984aa} thermostat with mass
$m_{\eta}=50.0$. For each temperature, 200 statistically independent
trajectories are equilibrated for $5\times10^{5}$ steps using a timestep
$\Delta\,t=0.001$. After equilibration, 50 configurations are sampled from
each trajectory every $10^{4}$ timesteps over a further $5\times10^{5}$ steps
giving a total set of $10^{4}$ configurations over which statistical analysis
of the interfacial properties is performed.

\subsection{Intrinsic Surface Analysis}
Analysis of the interfacial region at the shaper molecular scale requires an operational definition
of the collective coordinates of the intrinsic surface $\textstyle\xi\bb{\mathbf{x}}$ for a given molecular configuration.
This step is accomplished through a percolation analysis to separate the liquid phase -- identified by
the spanning cluster -- from the isolated clusters in the vapour phase.
Originally put forward by  Stillinger~\citep{Stillinger1982ab}, several approaches have been developed to this end
~\citep{Chacon2003aa,Tarazona2004aa,Chowdhary2008aa,Jorge2007aa,Partay2008ab,Jorge2010ab,Sega2013aa,Willard2010aa}.
In particular, the intrinsic sampling method (ISM) was the first implementation allowing a
quantitative comparison between computer simulations and CWT~\citep{Chacon2003aa,Tarazona2004aa}
and is the method of choice in the current analysis.
The ISM  can be interpreted as an outlier detection method~\citep{Rousseeuw2005aa} that iteratively estimates the amplitudes of a
Fourier series representation of the intrinsic surface
\begin{equation}
\label{eqn:meth2}
\xi(\mathbf{x};k_{u})=\lsum{\bbm{\mathbf{k}}\leq k_{u}}{}\,\hat{\xi}_{\mathbf{k}}\,e^{i\,\mathbf{k}\cdot\mathbf{x}}\;,
\end{equation}
\noindent where $\mathbf{x}=(x,y)$ is a vector parallel to the surface,  $\hat{\xi}_{k}$ are the amplitudes associated
with each wavevector in the basis set, $\mathbf{k}=2\pi(n_{x}/L_{x},n_{y}/L_{y})$,
with $n_{x},n_{y}=0,\pm1,\pm2,\dots$ and $k_{u}=2\pi/\lambda_{u}$ is an upper wave vector
cutoff that sets the lower resolution limit $\lambda_{u}$ of the surface.
Here the judicious choice within CWT is to set a cutoff wavelength commensurate
with the molecular size $\lambda_{u}\approx\sigma$~\citep{Chacon2003aa}.
A basic assumption behind ISM is that the molecular system consists of a set of surface atoms,
or \textit{inliers}, whose position is described by the assumed mathematical model, \eqn{eqn:meth2},
and \textit{outliers} -- representing the atoms in the liquid and vapour phases --
which are not accounted by the model.
The spanning cluster is obtained by a graph traversal algorithm that identifies the largest set of atoms with at
least $3$ neighbours at distances below $1.4$. From this set, an initial estimate of the surface \textit{inliers}
is obtained by dividing the $xy-$plane into a $3\times3$ mesh and finding the  $N_{s}=9$ atoms with
the largest $z$ coordinate in each mesh element.
The intrinsic surface is then defined as the surface of least area passing through the selected atoms.
This is accomplished by minimizing the objective function
\begin{equation}
\label{eqn:meth3}
\min\limits_{\bbm{\mathbf{k}}\leq k_{u}}f\bb{\hat{\xi}_{\mathbf{k}}}_{}\equiv
\frac{\omega}{2}\lsum{i=1}{N_{s}}\bb{z_{i} - \xi\bb{\mathbf{x}_{i}}}^{2} +
\frac{A}{2}\lsum{\bbm{\mathbf{k}}\leq k_{u}}{}\,\mathbf{k}^{2}\bbm{\hat{\xi}_{\mathbf{k}}}^{2}\;,
\end{equation}
\noindent where the $A=L_{x}L_{y}$ and $\omega$ is a constant that specifies the strength of the
harmonic restraints tethering the surface to the atomic positions.
The method then searches the percolating cluster for the closest atom to the minimal surface that
is not part of the set of \textit{inliers}.
The procedure is iterated until an optimum value of the surface density $n_{s}=N_{s}/A$  is reached.
At the highest level of resolution, the surface atoms are represented in the density profile
as a delta function $n_{s}\delta(z)$.
Too low a value of $n_{s}$ results in a unphysical shoulder between
the surface and the first layer in the liquid phase of the intrinsic density profile $\tilde{\rho}\bb{z}$ and too large
values are characterised by a  degenerate mathematical surface that tries to fit all the surface atoms~\citep{Chacon2009aa}.
For the LJ system at the studied temperatures, it was found $n_{s}=0.8$ to best describe the interface at the molecular scale
with a clear peak at the origin followed by a significant oscillatory structure whose amplitude decays towards the liquid bulk density.

%%%%%%%%%%%%%%%%%%%%%%%%
\section{Results}
\label{sec:results}
%% snapshot of an intrinsic interface
%\begin{figure} [H]      % placement: Here, here, top, bottom or page
%\centering
%\includegraphics[width=0.9\linewidth, angle=0.0]{figures/fig_snapshot}
%\caption{Graphical representation of a liquid-vapour molecular configuration.
%The liquid phase (blue atoms) is separated by the vapour phase (grey atoms)
%by a surface (yellow colour) that passes through the surface atoms (red colour) according to \eqn{eqn:meth2} and \eqn{eqn:meth3}.}
%\label{fig:snapshot}
%\end{figure}
% intrinsic density profiles
%
\begin{figure}     % placement: Here, here, top, bottom or page
\centering
\includegraphics[width=0.9\linewidth, angle=0.0]{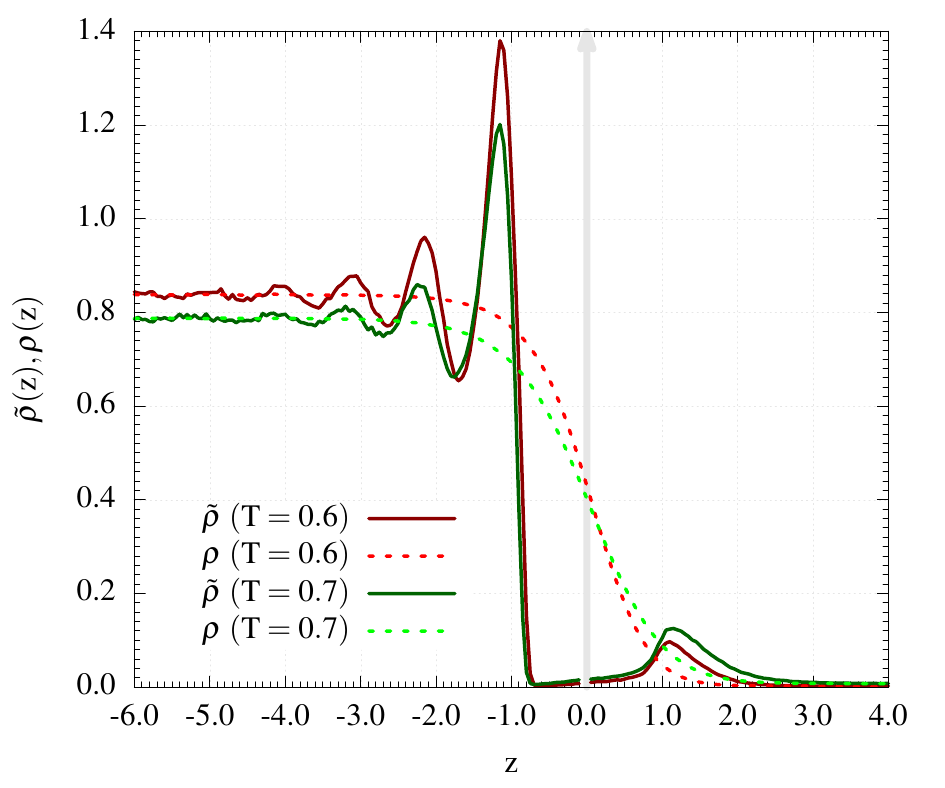}
\caption{(Colour online). Intrinsic density profiles $\tilde{\rho}(z)$ (solid line) compared to mean density $\rho(z)$ (dashed line)
for systems at $T=0.6$ (red colour) and $T=0.7$ (green colour), respectively.
Surface atoms are represented by the delta function, represented as a grey arrow at $z=0$.
The mean density profiles were centred at their corresponding Gibbs dividing surfaces $\bba{\xi}$.}
\label{fig:density}
\end{figure}
\fig{fig:density} depicts the intrinsic profiles at $T=0.6$ and $T=0.7$
respectively, obtained using the largest wavevector cutoff $k_{u}=2\pi$.
Intrinsic quantities are plotted in a frame centred on the intrinsic surface,
with the corresponding fixed grid quantities centred on the location of the
appropriate Gibbs dividing surface. For convenience, these are both denoted
as being at $z=0$ in all Figures. The removal of the blurring effect of
thermal fluctuations results in the clear appearance of the molecular
structure at the interface, given by the delta-function at $z=0$ representing
the surface layer, followed by a strong oscillatory structure whose amplitude
decays to zero at the bulk liquid density, and an adsorption peak at
$z\approx1.0$ in agreement with previous
results~\citep{Chacon2009aa,Chacon2005aa}.

\begin{figure}
\centering
\includegraphics[width=1.0\linewidth, angle=0.0]{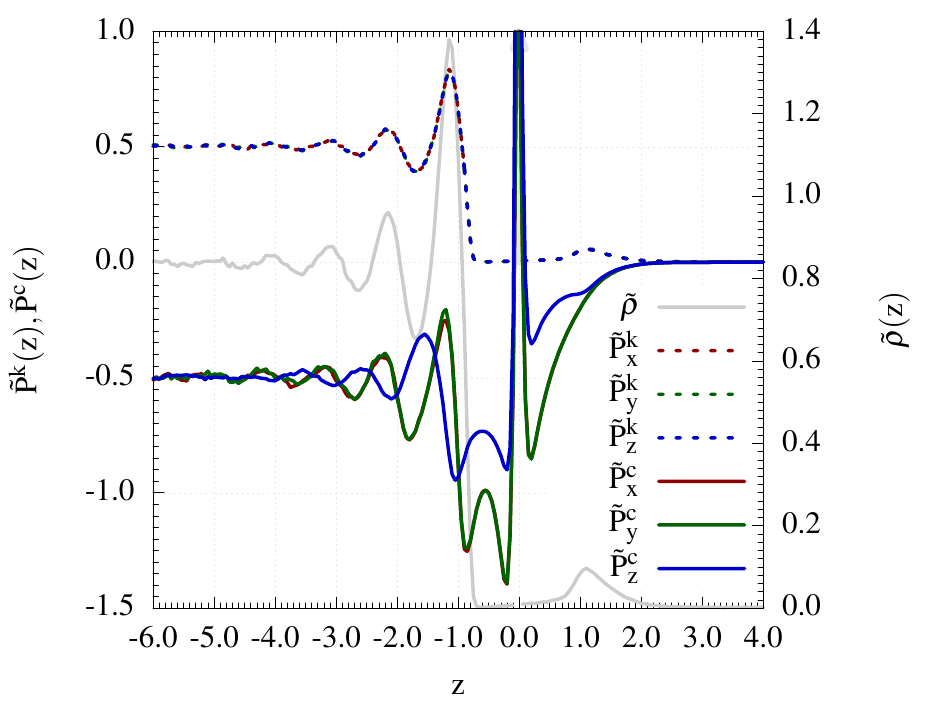}
\caption{(Colour online). Kinetic and configurational components of the
intrinsic pressure tensor, $\tilde{\mathbf{P}}^{k}\bb{z}$ and  $\tilde{\mathbf{P}}^{c}\bb{z}$, at $T=0.6$.
At constant temperature,
$\tilde{P}^{k}_{x}$ (dashed red), $\tilde{P}^{k}_{y}$ (dashed green) and $\tilde{P}^{k}_{z}$ (dashed blue), are
identical and proportional to the intrinsic density.
Cylindrical symmetry about the normal axis gives identical configurational terms parallel to the interface,
$\tilde{P}^{c}_{x}$ (red line) and $\tilde{P}^{c}_{y}$ (green line), within statistical uncertainty.
The configurational pressure normal to the interface, $\tilde{P}^{c}_{c}$ (blue line), is
significantly larger in the intermediate regions between the peaks the intrinsic density profile.
The intrinsic density, $\tilde{\rho}\bb{z}$ (grey line), is plotted on the secondary vertical axis.}
\label{fig:fig_pressure_1}
\end{figure}

The kinetic and configurational components of the intrinsic pressure profile
(cf. \eqn{eqn:pres3}) at $T=0.6$ are depicted in \fig{fig:fig_pressure_1}.
The dyadic terms $m_{i}\mathbf{v}_{i}\mathbf{v}_{i}$ in the kinetic component
$\tilde{\mathbf{P}}^{k}$, being a single particle property, are as result
directly proportional to the intrinsic density through the kinetic equation
of state $\tilde{P}^{k}\bb{z}=\tilde{\rho}\bb{z}\,k_{B}T$, where $k_{B}$ is
the Boltzmann's constant. At constant temperature, these are all equal,
emphasising the well known fact that, in the absence of temperature
gradients, surface tension is a configurational property as attested by the
configurational terms $\tilde{\mathbf{P}}^{c}$~\citep{Kirkwood1949aa}. On the
liquid phase, away from the interface (viz. $\textstyle z\lesssim-4.0$),
these are equivalent by a symmetry argument, i.e. the configurational
pressure tensor is isotropic (the negative value reflects the cohesive energy
of the LJ interactions).

% kinetic and configurational components of the pressure tensor
%\begin{figure}
%\centering
%\begin{subfigure}[b]{1.0\linewidth}
%\includegraphics[width=1.0\linewidth]{figures/fig_press_T0_6_kin}
%%\caption{}
%\end{subfigure}
%\begin{subfigure}[b]{1.0\linewidth}
%\includegraphics[width=1.0\linewidth]{figures/fig_press_T0_6_con}
%%\caption{}
%\end{subfigure}
%\caption{(Colour online).  Top panel: kinetic components of the intrinsic pressure tensor, $\tilde{\mathbf{P}}^{k}\bb{z}$, at $T=0.6$.
%At constant temperature,
%$\tilde{P}^{k}_{x}$ (red), $\tilde{P}^{k}_{y}$ (green) and $\tilde{P}^{k}_{z}$ (blue), are
%identical and proportional to the intrinsic density through the kinetic relation $\tilde{P}^{k}=\tilde{\rho}\,T$.
%The intrinsic density, $\tilde{\rho}\bb{z}$ (grey), is plotted on the secondary vertical axis.
%Bottom panel: configurational components of the intrinsic pressure tensor, $\tilde{\mathbf{P}}^{c}\bb{z}$ , at $T=0.6$.
%Cylindrical symmetry about the normal axis gives identical components parallel to the interface,
%$\tilde{P}^{c}_{x}$ (red) and $\tilde{P}^{c}_{y}$ (green), within statistical uncertainty.
%The configurational pressure normal to the interface, $\tilde{P}^{c}_{c}$ (blue), is
%significantly larger in the regions between the maxima in the intrinsic density (same notation as top panel),
%resulting in an excess pressure. The results at $T=0.7$ exhibit similar behaviour.}
%\label{fig:fig_pressure_1}
%\end{figure}

\begin{figure}
\centering
\begin{subfigure}[b]{1.0\linewidth}
\includegraphics[width=1.0\linewidth]{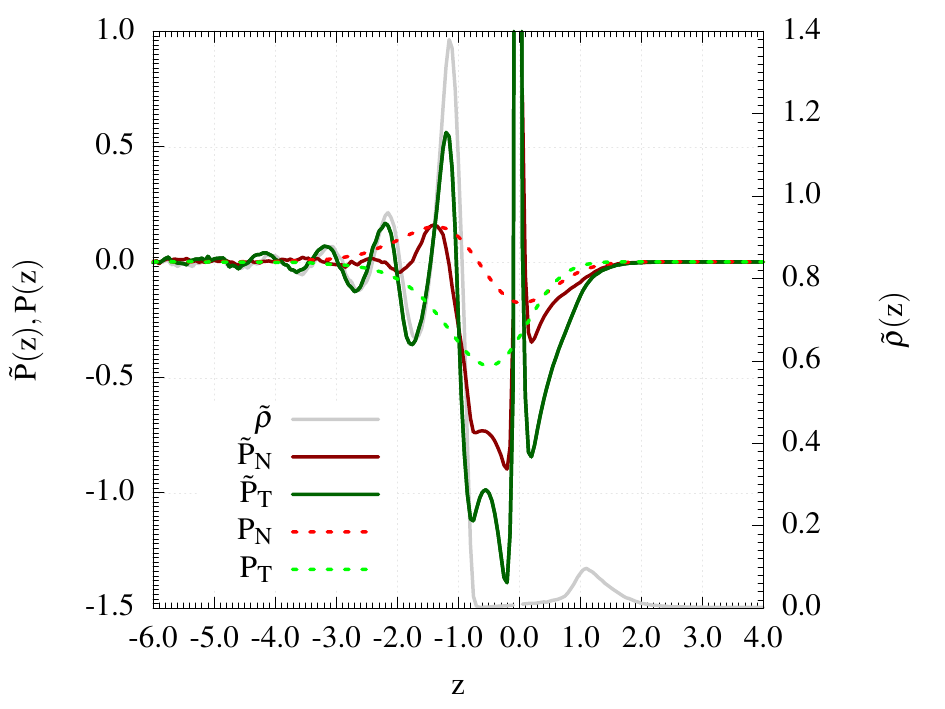}
%\caption{}
\end{subfigure}
\begin{subfigure}[b]{1.0\linewidth}
\includegraphics[width=1.0\linewidth]{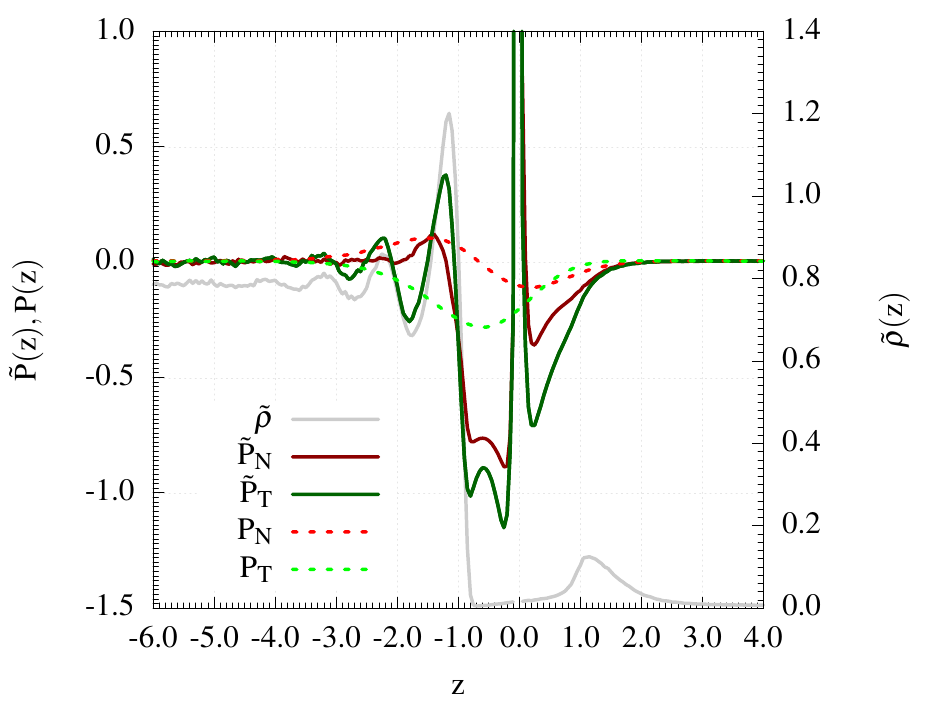}
%\caption{}
\end{subfigure}
\caption{(Colour online). Top panel: components of the intrinsic pressure tensor
tensor normal, $\tilde{P}_{N}=\tilde{P}_{z}$ (red colour),
and transversal to the interface, $\tilde{P}_{T}=(\tilde{P}_{x}+\tilde{P}_{y})/2$ (green colour), at $T=0.6$.
The mean profiles $P_{N}$ (dashed red line) and $P_{T}$ (dashed green line) are plotted for comparison.
The intrinsic density, $\tilde{\rho}\bb{z}$ (grey), is plotted on the secondary vertical axis.
Bottom panel: intrinsic and average pressure profiles normal and transversal to the interface at $T=0.7$. The notation is the same as top panel.}
\label{fig:fig_pressure_2}
\end{figure}

In the layers closer to the interface, fluctuations in density break the symmetry of interactions.
\fig{fig:fig_pressure_2} illustrates this difference in more detail for $\tilde{P}_{N}=\tilde{P}_{z}$ and
$\tilde{P}_{T}=(\tilde{P}_{x}+\tilde{P}_{y})/2$, at $T=0.6$ and $T=0.7$ respectively.
The transverse pressure $\tilde{P}_{T}$ follows closely the oscillations in density,
supporting the notion that, at the peaks in the intrinsic density, the larger number of neighbours
interacting with an atom in the directions parallel to the interface results
in a higher proportion of short ranged repulsive interactions contributing to the extra stress.

% comparison between transverse and normal components of the
% intrinsic pressure and mean pressure profiles at T=0.6 and T=0.7

On the other hand, in the normal direction, $\tilde{P}_{N}$ exhibits a more steady profile
up until the region between the first two layers in the liquid phase.
While being zero at the peak locations, reflecting mechanical equilibrium of the liquid layers
(cf.~\fig{fig:fig_pressure_2}, viz.~$\textstyle z\approx -1\times2^{1/6}$ and $\textstyle z\approx -2\times2^{1/6}$),
there is a clear positive excess normal pressure between these at $z\approx 1.5$.
The physical argument behind this excess energy resides in the observation that an atom located at this position
would experience the short range repulsive forces of the adjacent fluid layers.
The resultant instability would force the atom to transit to one of the layers thus preserving the liquid structure
next to the interface.
We note that at larger distances from the interface, the mean normal pressure $P_N$ is equal in the liquid and vapour phase, as can be seen in \fig{fig:fig_pressure_2}. Over the surface the mean divergence of pressure will still be equal to zero, i.e. mechanical equilibrium must be satisfied as the surface is stationary.

% comparison between the excess pressure and surface tension integral
% and corresponding mean profiles at T=0.6 and T=0.7 respectively
\begin{figure}
\centering
\begin{subfigure}[b]{1.0\linewidth}
\includegraphics[width=1.0\linewidth]{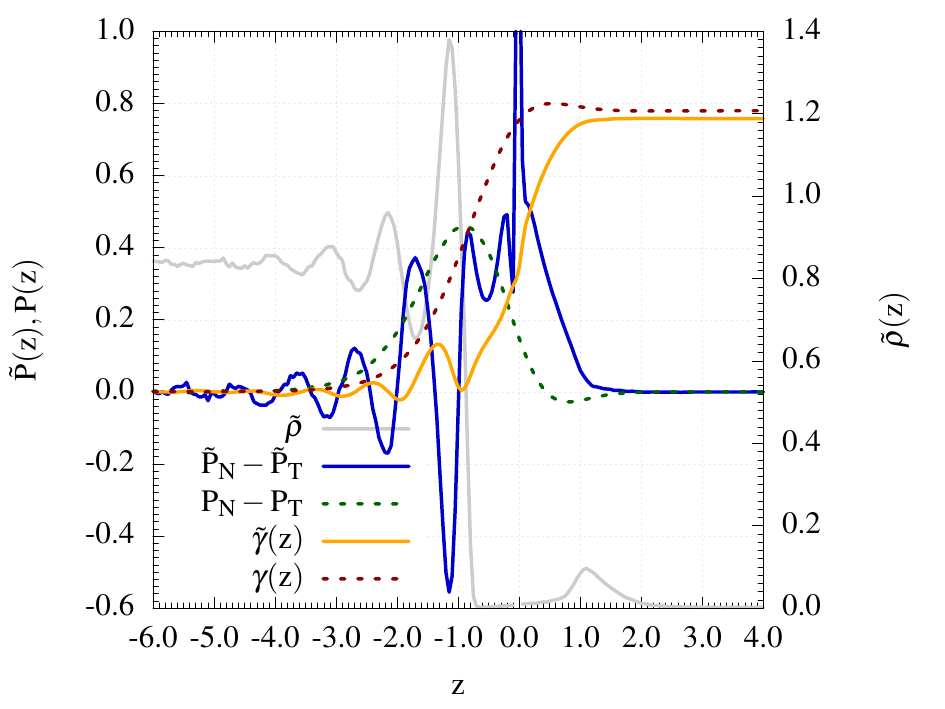}
%\caption{}
\end{subfigure}
\begin{subfigure}[b]{1.0\linewidth}
\includegraphics[width=1.0\linewidth]{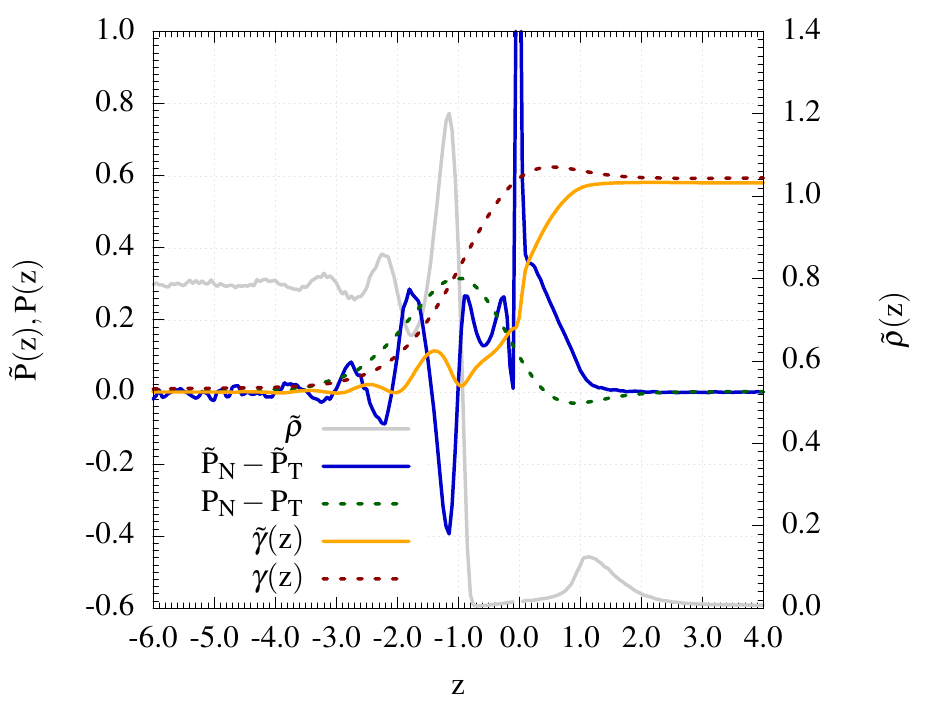}
%\caption{}
\end{subfigure}
\caption{(Colour online). Top panel: comparison between excess intrinsic pressure,
$\tilde{P}_{N} - \tilde{P}_{T}$ (blue colour), and its mean counterpart (green colour) at $T=0.6$.
The bottom panel shows the results at $T=0.7$. The notation is the same as top panel.}
\label{fig:fig_pressure_gamma}
\end{figure}

\fig{fig:fig_pressure_gamma} depicts the excess pressure of the fluid,
defined by $\tilde{P}_{N}\bb{z} - \tilde{P}_{T}\bb{z}$, whose integral over
the entire domain, $\textstyle \gamma\bb{z} =
\int_{-\infty}^{z}\dd{z^{\prime}}\bb{P_{N}\bb{z^{\prime}} -
P_{T}\bb{z^{\prime}}}$, gives the Kirkwood-Buff~\citep{Kirkwood1951aa}
definition of the surface tension (cf.~\eqn{eqn:tension2}
and~\eqn{eqn:tension3}). At the highest molecular resolution, the intrinsic
pressure tensor \eqn{eqn:pres3} allows us to ascertain the microscopic
mechanism involved in giving rise to the interfacial energy.

We can observe that the surface tension has contributions from both the
liquid and vapour facing sides of the (intrinsic) interface.
Importantly, however, we note that the depletion region between the
first layer of fluid at the surface layer is where the onset of the surface
tension takes place. We are also able to see that the adsorption peak --
approximately one fluid layer away from the surface -- is vital to the
contribution of the other half of the surface tension. Crucially, this is
truly an adsorption peak rather than a density oscillation around the bulk
vapour density (in contrast to the liquid facing side of the interface), due
to the attraction of the vapour particles to the liquid (of course, as they
are the same atoms). As the density increases and returns to the bulk
vapour density, there is a net additive contribution to surface tension.

The physical picture is thus as follows: the surface tension, unlike the
traditional mechanical picture -- defined as the excess pressure acting
across a surface of zero width between the liquid and vapour phases --
arises instead across a boundary region spanning the three key layers involved
in the intrinsic surface -- the adsorbed layer at $z\approx 1.0$,
the surface layer at $z=0$ and the first layer in the liquid phase at
$z\approx -1.0$. This region is constructed by pairs of layers
attracting each other in an interleaved manner, at a distance commensurate
with the attractive tail of the Lennard-Jones potential -- the adsorbed layer
at $z\approx 1.0$ with the first liquid layer at $z\approx-1.0$ and
surface layer at $z=0$ with the second liquid layer at $z\approx-2.0$ respectively.

The results shown in \fig{fig:fig_pressure_gamma}
demonstrate that the surface tension force has similar contributions
on either side of the interface, an observation first given
in~\citep{Sega2015aa}, but shown here with unprecedented clarity using the
intrinsic pressure derived in this work.

%%%%%%%%%%%%%%%%%%%%%%%%%%%%%%%%
% Conclusions
\section{Conclusion}
We have proposed a generalisation of Irving and Kirkwood's expression for the
pressure tensor free from the smearing effect of thermal fluctuations of
capillary waves. This is accomplished by letting the contour of integration
become a dynamic quantity, unique to each pair of interacting particles, that
reflects the instantaneous local curvature of the interface. Expressing
hydrodynamic fields in control volume format, we are able to compute the
intrinsic density profile that matches the profiles reported previously.

Applying the same technique to pressure allows the derivation of an intrinsic pressure tensor,
which gives unprecedented insight into the liquid-vapour interface.
More importantly, however, the notion of a mathematical surface of infinitesimal width,
central to the mechanical definition of the surface tension, looses meaning at the
microscopic level where the discrete nature of matter becomes apparent.
Instead, we observe the emergence of the surface tension over a finite range of three atomic layers,
acting exactly across the intrinsic surface.

The surface tension and other quantities of interest are derived from the
continuity equations of hydrodynamics -- expressed in a Lagrangian frame of
reference relative to the intrinsic surface -- and therefore are valid
arbitrarily far from equilibrium. This makes the derived expressions
particularly useful to the study of non equilibrium inhomogeneous systems --
e.g., Marangoni flow, hydrodynamic instabilities and bubble nucleation, which
are inherently non-local and unsteady. Of particular interest would also be
extension of this study to systems in confinement including heterogeneous
substrates and nanochannels (e.g.~Refs~\onlinecite{Peter2017,Matteo2017}).

%In this work we have applied the density and stress calculation of \citep{Kirkwood1949aa} in volumes which move relative to an intrinsic interface \citep{Rowlinson2002aa}. This provides an a-priori derivation of the instantaneous expressions for density; which when averaged match the intrinsic density postulated by capillary wave theory \citep{Chacon2001aa}. Applying the same technique to pressure allows the derivation of an intrinsic pressure tensor, which gives unprecedented insight into the liquid-vapour interface. This pressure is used in the \citep{Kirkwood1951aa} formula to give a surface tension calculation relative to the moving interface. A number of results are presented for a Lennard-Jones fluid, which show that kinetic pressure does not contribute to surface tension and that contributions to the surface tension are localised to within $1.0$ reduced units of the surface. The resulting expressions are purely mechanical and therefore valid arbitrarily far from equilibrium. This validity is 
essential in a range of important industrial processes where the surface tension plays a role, such as Marangoni flows, hydrodynamic instabilities and bubble nucleation, which are inherently non-equilibrium, non-local and unsteady.

%%%%%%%%%%%%%%%%%%%%%%%%%%%%%%%%
% Tail of the paper
\begin{acknowledgments}
We acknowledge financial support from the European Research Council (ERC)
through Advanced Grant No. 247031 and by the Engineering and Physical
Sciences Research Council (EPSRC) of the UK through Grant No. EP/L020564.
\end{acknowledgments}

% Appendix
%\input{Appendix}
%\bibliographystyle{phaip}
%\bibliography{IntrinsicPressureTensor}

\begin{thebibliography}{51}%
\makeatletter
\providecommand \@ifxundefined [1]{%
 \@ifx{#1\undefined}
}%
\providecommand \@ifnum [1]{%
 \ifnum #1\expandafter \@firstoftwo
 \else \expandafter \@secondoftwo
 \fi
}%
\providecommand \@ifx [1]{%
 \ifx #1\expandafter \@firstoftwo
 \else \expandafter \@secondoftwo
 \fi
}%
\providecommand \natexlab [1]{#1}%
\providecommand \enquote  [1]{``#1''}%
\providecommand \bibnamefont  [1]{#1}%
\providecommand \bibfnamefont [1]{#1}%
\providecommand \citenamefont [1]{#1}%
\providecommand \href@noop [0]{\@secondoftwo}%
\providecommand \href [0]{\begingroup \@sanitize@url \@href}%
\providecommand \@href[1]{\@@startlink{#1}\@@href}%
\providecommand \@@href[1]{\endgroup#1\@@endlink}%
\providecommand \@sanitize@url [0]{\catcode `\\12\catcode `\$12\catcode
  `\&12\catcode `\#12\catcode `\^12\catcode `\_12\catcode `\%12\relax}%
\providecommand \@@startlink[1]{}%
\providecommand \@@endlink[0]{}%
\providecommand \url  [0]{\begingroup\@sanitize@url \@url }%
\providecommand \@url [1]{\endgroup\@href {#1}{\urlprefix }}%
\providecommand \urlprefix  [0]{URL }%
\providecommand \Eprint [0]{\href }%
\providecommand \doibase [0]{http://dx.doi.org/}%
\providecommand \selectlanguage [0]{\@gobble}%
\providecommand \bibinfo  [0]{\@secondoftwo}%
\providecommand \bibfield  [0]{\@secondoftwo}%
\providecommand \translation [1]{[#1]}%
\providecommand \BibitemOpen [0]{}%
\providecommand \bibitemStop [0]{}%
\providecommand \bibitemNoStop [0]{.\EOS\space}%
\providecommand \EOS [0]{\spacefactor3000\relax}%
\providecommand \BibitemShut  [1]{\csname bibitem#1\endcsname}%
\let\auto@bib@innerbib\@empty
%</preamble>
\bibitem [{\citenamefont {Rowlinson}\ and\ \citenamefont
  {Widom}(2002)}]{Rowlinson2002aa}%
  \BibitemOpen
  \bibfield  {author} {\bibinfo {author} {\bibfnamefont {J.~S.}\ \bibnamefont
  {Rowlinson}}\ and\ \bibinfo {author} {\bibfnamefont {B.}~\bibnamefont
  {Widom}},\ }\href {http://books.google.co.uk/books?id={\_}ydSF{\_}XUVeEC}
  {\emph {\bibinfo {title} {{Molecular Theory of Capillarity}}}},\ Dover books
  on chemistry\ (\bibinfo  {publisher} {Dover Publications},\ \bibinfo {year}
  {2002})\BibitemShut {NoStop}%
\bibitem [{\citenamefont {Sanyal}\ \emph {et~al.}(1991)\citenamefont {Sanyal},
  \citenamefont {Sinha}, \citenamefont {Huang},\ and\ \citenamefont
  {Ocko}}]{Sanyal1991aa}%
  \BibitemOpen
  \bibfield  {author} {\bibinfo {author} {\bibfnamefont {M.~K.}\ \bibnamefont
  {Sanyal}}, \bibinfo {author} {\bibfnamefont {S.~K.}\ \bibnamefont {Sinha}},
  \bibinfo {author} {\bibfnamefont {K.~G.}\ \bibnamefont {Huang}}, \ and\
  \bibinfo {author} {\bibfnamefont {B.~M.}\ \bibnamefont {Ocko}},\ }\bibfield
  {title} {\enquote {\bibinfo {title} {{X-ray-scattering study of
  capillary-wave fluctuations at a liquid surface}},}\ }\href {\doibase
  10.1103/PhysRevLett.66.628} {\bibfield  {journal} {\bibinfo  {journal}
  {Physical Review Letters}\ }\textbf {\bibinfo {volume} {66}},\ \bibinfo
  {pages} {628--631} (\bibinfo {year} {1991})}\BibitemShut {NoStop}%
\bibitem [{\citenamefont {Tidswell}\ \emph {et~al.}(1991)\citenamefont
  {Tidswell}, \citenamefont {Rabedeau}, \citenamefont {Pershan},\ and\
  \citenamefont {Kosowsky}}]{Tidswell1991aa}%
  \BibitemOpen
  \bibfield  {author} {\bibinfo {author} {\bibfnamefont {I.~M.}\ \bibnamefont
  {Tidswell}}, \bibinfo {author} {\bibfnamefont {T.~a.}\ \bibnamefont
  {Rabedeau}}, \bibinfo {author} {\bibfnamefont {P.~S.}\ \bibnamefont
  {Pershan}}, \ and\ \bibinfo {author} {\bibfnamefont {S.~D.}\ \bibnamefont
  {Kosowsky}},\ }\bibfield  {title} {\enquote {\bibinfo {title} {{Complete
  wetting of a rough surface: An x-ray study}},}\ }\href {\doibase
  10.1103/PhysRevLett.66.2108} {\bibfield  {journal} {\bibinfo  {journal}
  {Physical Review Letters}\ }\textbf {\bibinfo {volume} {66}},\ \bibinfo
  {pages} {2108--2111} (\bibinfo {year} {1991})}\BibitemShut {NoStop}%
\bibitem [{\citenamefont {Buff}\ \emph {et~al.}(1965)\citenamefont {Buff},
  \citenamefont {Lovett}, \citenamefont {Jr},\ and\ \citenamefont
  {Stillinger}}]{Buff1965ab}%
  \BibitemOpen
  \bibfield  {author} {\bibinfo {author} {\bibfnamefont {F.~P.}\ \bibnamefont
  {Buff}}, \bibinfo {author} {\bibfnamefont {R.~A.}\ \bibnamefont {Lovett}},
  \bibinfo {author} {\bibfnamefont {F.~S.}\ \bibnamefont {Jr}}, \ and\ \bibinfo
  {author} {\bibfnamefont {F.~H.}\ \bibnamefont {Stillinger}},\ }\bibfield
  {title} {\enquote {\bibinfo {title} {{Interfacial Density Profile for Fluids
  in the Critical Region}},}\ }\href {\doibase 10.1103/PhysRevLett.15.621}
  {\bibfield  {journal} {\bibinfo  {journal} {Physical Review Letters}\
  }\textbf {\bibinfo {volume} {15}},\ \bibinfo {pages} {621--623} (\bibinfo
  {year} {1965})}\BibitemShut {NoStop}%
\bibitem [{\citenamefont {Evans}(1979)}]{Evans1979aa}%
  \BibitemOpen
  \bibfield  {author} {\bibinfo {author} {\bibfnamefont {R.}~\bibnamefont
  {Evans}},\ }\bibfield  {title} {\enquote {\bibinfo {title} {{The nature of
  the liquid-vapour interface and other topics in the statistical mechanics of
  non-uniform, classical fluids}},}\ }\href {\doibase
  10.1080/00018737900101365} {\bibfield  {journal} {\bibinfo  {journal}
  {Advances in Physics}\ }\textbf {\bibinfo {volume} {28}},\ \bibinfo {pages}
  {143--200} (\bibinfo {year} {1979})}\BibitemShut {NoStop}%
\bibitem [{\citenamefont {Percus}(1986)}]{Percus1986aa}%
  \BibitemOpen
  \bibfield  {author} {\bibinfo {author} {\bibfnamefont {J.~R.}\ \bibnamefont
  {Percus}},\ }\href {https://books.google.co.uk/books?id=v6bvAAAAMAAJ} {\emph
  {\bibinfo {title} {{Fluid Interfacial Phenomena}}}},\ edited by\ \bibinfo
  {editor} {\bibfnamefont {C.~A.}\ \bibnamefont {Croxton}},\ Wiley-Interscience
  publication\ (\bibinfo  {publisher} {Wiley},\ \bibinfo {year}
  {1986})\BibitemShut {NoStop}%
\bibitem [{\citenamefont {Fern{\'{a}}ndez}\ \emph {et~al.}(2013)\citenamefont
  {Fern{\'{a}}ndez}, \citenamefont {Chac{\'{o}}n}, \citenamefont {Tarazona},
  \citenamefont {Parry},\ and\ \citenamefont {Rasc{\'{o}}n}}]{Fernandez2013aa}%
  \BibitemOpen
  \bibfield  {author} {\bibinfo {author} {\bibfnamefont {E.~M.}\ \bibnamefont
  {Fern{\'{a}}ndez}}, \bibinfo {author} {\bibfnamefont {E.}~\bibnamefont
  {Chac{\'{o}}n}}, \bibinfo {author} {\bibfnamefont {P.}~\bibnamefont
  {Tarazona}}, \bibinfo {author} {\bibfnamefont {A.~O.}\ \bibnamefont {Parry}},
  \ and\ \bibinfo {author} {\bibfnamefont {C.}~\bibnamefont {Rasc{\'{o}}n}},\
  }\bibfield  {title} {\enquote {\bibinfo {title} {{Intrinsic Fluid Interfaces
  and Nonlocality}},}\ }\href {\doibase 10.1103/PhysRevLett.111.096104}
  {\bibfield  {journal} {\bibinfo  {journal} {Physical Review Letters}\
  }\textbf {\bibinfo {volume} {111}},\ \bibinfo {pages} {096104} (\bibinfo
  {year} {2013})}\BibitemShut {NoStop}%
\bibitem [{\citenamefont {M{\"{u}}ller}\ and\ \citenamefont
  {M{\"{u}}nster}(2005)}]{Muller2005aa}%
  \BibitemOpen
  \bibfield  {author} {\bibinfo {author} {\bibfnamefont {M.}~\bibnamefont
  {M{\"{u}}ller}}\ and\ \bibinfo {author} {\bibfnamefont {G.}~\bibnamefont
  {M{\"{u}}nster}},\ }\bibfield  {title} {\enquote {\bibinfo {title} {{Profile
  and width of rough interfaces}},}\ }\href {\doibase
  10.1007/s10955-004-8824-2} {\bibfield  {journal} {\bibinfo  {journal}
  {Journal of Statistical Physics}\ }\textbf {\bibinfo {volume} {118}},\
  \bibinfo {pages} {669--686} (\bibinfo {year} {2005})}\BibitemShut {NoStop}%
\bibitem [{Die(1999)}]{Dietrich1999}%
  \BibitemOpen
  \bibfield  {title} {\enquote {\bibinfo {title} {Effective hamiltonian for
  liquid-vapour interfaces},}\ }\href
  {https://journals.aps.org/pre/abstract/10.1103/PhysRevE.59.6766} {\bibfield
  {journal} {\bibinfo  {journal} {Phys. Rev. E}\ }\textbf {\bibinfo {volume}
  {59}},\ \bibinfo {pages} {6766--6784} (\bibinfo {year} {1999})}\BibitemShut
  {NoStop}%
\bibitem [{Par(2000)}]{Parry2000}%
  \BibitemOpen
  \bibfield  {title} {\enquote {\bibinfo {title} {Critical effects at 3d wedge
  filling},}\ }\href
  {https://journals.aps.org/prl/abstract/10.1103/PhysRevLett.85.345} {\bibfield
   {journal} {\bibinfo  {journal} {Phys. Rev. Lett.}\ }\textbf {\bibinfo
  {volume} {85}},\ \bibinfo {pages} {345} (\bibinfo {year} {2000})}\BibitemShut
  {NoStop}%
\bibitem [{\citenamefont {H$\ddot{\rm o}$fling}\ and\ \citenamefont
  {Dietrich}(2015)}]{Dietrich2015}%
  \BibitemOpen
  \bibfield  {author} {\bibinfo {author} {\bibfnamefont {F.}~\bibnamefont
  {H$\ddot{\rm o}$fling}}\ and\ \bibinfo {author} {\bibfnamefont
  {S.}~\bibnamefont {Dietrich}},\ }\bibfield  {title} {\enquote {\bibinfo
  {title} {Enhanced wavelength-dependent surface tension of liquid-vapour
  interfaces},}\ }\href
  {http://iopscience.iop.org/article/10.1209/0295-5075/109/46002/meta}
  {\bibfield  {journal} {\bibinfo  {journal} {Europhys. Lett.}\ }\textbf
  {\bibinfo {volume} {109}},\ \bibinfo {pages} {46002} (\bibinfo {year}
  {2015})}\BibitemShut {NoStop}%
\bibitem [{\citenamefont {Yatsyshin}, \citenamefont {Parry},\ and\
  \citenamefont {Kalliadasis}(2016)}]{Peter2016}%
  \BibitemOpen
  \bibfield  {author} {\bibinfo {author} {\bibfnamefont {P.}~\bibnamefont
  {Yatsyshin}}, \bibinfo {author} {\bibfnamefont {A.~O.}~\bibnamefont {Parry}}, \
  and\ \bibinfo {author} {\bibfnamefont {S.}~\bibnamefont {Kalliadasis}},\
  }\bibfield  {title} {\enquote {\bibinfo {title} {Complete prewetting},}\
  }\href {http://iopscience.iop.org/article/10.1088/0953-8984/28/27/275001}
  {\bibfield  {journal} {\bibinfo  {journal} {J. Phys.: Condens. Matter}\
  }\textbf {\bibinfo {volume} {28}},\ \bibinfo {pages} {267001} (\bibinfo
  {year} {2016})}\BibitemShut {NoStop}%
\bibitem [{\citenamefont {Sega}, \citenamefont {F{\'{a}}bi{\'{a}}n},\ and\
  \citenamefont {Jedlovszky}(2015)}]{Sega2015aa}%
  \BibitemOpen
  \bibfield  {author} {\bibinfo {author} {\bibfnamefont {M.}~\bibnamefont
  {Sega}}, \bibinfo {author} {\bibfnamefont {B.}~\bibnamefont
  {F{\'{a}}bi{\'{a}}n}}, \ and\ \bibinfo {author} {\bibfnamefont
  {P.}~\bibnamefont {Jedlovszky}},\ }\bibfield  {title} {\enquote {\bibinfo
  {title} {{Layer-by-layer and intrinsic analysis of molecular and
  thermodynamic properties across soft interfaces Layer-by-layer and intrinsic
  analysis of molecular and thermodynamic properties across soft
  interfaces}},}\ }\href {\doibase 10.1063/1.4931180} {\bibfield  {journal}
  {\bibinfo  {journal} {The Journal of Chemical Physics}\ }\textbf {\bibinfo
  {volume} {114709}} (\bibinfo {year} {2015}),\ 10.1063/1.4931180}\BibitemShut
  {NoStop}%
\bibitem [{\citenamefont {Sega}\ \emph {et~al.}(2016)\citenamefont {Sega},
  \citenamefont {F{\'{a}}bi{\'{a}}n}, \citenamefont {Horvai},\ and\
  \citenamefont {Jedlovszky}}]{Sega2016ab}%
  \BibitemOpen
  \bibfield  {author} {\bibinfo {author} {\bibfnamefont {M.}~\bibnamefont
  {Sega}}, \bibinfo {author} {\bibfnamefont {B.}~\bibnamefont
  {F{\'{a}}bi{\'{a}}n}}, \bibinfo {author} {\bibfnamefont {G.}~\bibnamefont
  {Horvai}}, \ and\ \bibinfo {author} {\bibfnamefont {P.}~\bibnamefont
  {Jedlovszky}},\ }\bibfield  {title} {\enquote {\bibinfo {title} {{How is the
  surface tension of various liquids distributed along the interface
  normal?}}}\ }\href {\doibase 10.1021/acs.jpcc.6b09880} {\bibfield  {journal}
  {\bibinfo  {journal} {Journal of Physical Chemistry C}\ }\textbf {\bibinfo
  {volume} {120}},\ \bibinfo {pages} {27468--27477} (\bibinfo {year}
  {2016})}\BibitemShut {NoStop}%
\bibitem [{\citenamefont {Todd}, \citenamefont {Evans},\ and\ \citenamefont
  {Daivis}(1995)}]{Todd1995aa}%
  \BibitemOpen
  \bibfield  {author} {\bibinfo {author} {\bibfnamefont {B.~D.~B.}\
  \bibnamefont {Todd}}, \bibinfo {author} {\bibfnamefont {D.~D.~J.}\
  \bibnamefont {Evans}}, \ and\ \bibinfo {author} {\bibfnamefont {P.~P.~J.}\
  \bibnamefont {Daivis}},\ }\bibfield  {title} {\enquote {\bibinfo {title}
  {{Pressure tensor for inhomogeneous fluids}},}\ }\href {\doibase
  10.1103/PhysRevE.52.1627} {\bibfield  {journal} {\bibinfo  {journal}
  {Physical Review E}\ }\textbf {\bibinfo {volume} {52}},\ \bibinfo {pages}
  {1627} (\bibinfo {year} {1995})}\BibitemShut {NoStop}%
\bibitem [{\citenamefont {Baus}\ and\ \citenamefont
  {Lovett}(1990)}]{Baus1990aa}%
  \BibitemOpen
  \bibfield  {author} {\bibinfo {author} {\bibfnamefont {M.}~\bibnamefont
  {Baus}}\ and\ \bibinfo {author} {\bibfnamefont {R.}~\bibnamefont {Lovett}},\
  }\bibfield  {title} {\enquote {\bibinfo {title} {{Generalization of the
  stress tensor to nonuniform fluids and solids and its relation to
  Saint-Venants strain compatibility conditions}},}\ }\href {\doibase
  10.1103/PhysRevLett.65.1781} {\bibfield  {journal} {\bibinfo  {journal}
  {Physical Review Letters}\ }\textbf {\bibinfo {volume} {65}},\ \bibinfo
  {pages} {1781--1783} (\bibinfo {year} {1990})}\BibitemShut {NoStop}%
\bibitem [{\citenamefont {Irving}\ and\ \citenamefont
  {Kirkwood}(1950)}]{Irving1950aa}%
  \BibitemOpen
  \bibfield  {author} {\bibinfo {author} {\bibfnamefont {J.~H.}\ \bibnamefont
  {Irving}}\ and\ \bibinfo {author} {\bibfnamefont {J.~G.}\ \bibnamefont
  {Kirkwood}},\ }\bibfield  {title} {\enquote {\bibinfo {title} {{The
  Statistical Mechanical Theory of Transport Processes. IV. The Equations of
  Hydrodynamics}},}\ }\href {\doibase 10.1063/1.1747782} {\bibfield  {journal}
  {\bibinfo  {journal} {The Journal of Chemical Physics}\ }\textbf {\bibinfo
  {volume} {18}},\ \bibinfo {pages} {817} (\bibinfo {year} {1950})}\BibitemShut
  {NoStop}%
\bibitem [{\citenamefont {Harasima}(1958)}]{Harasima1958aa}%
  \BibitemOpen
  \bibfield  {author} {\bibinfo {author} {\bibfnamefont {A.}~\bibnamefont
  {Harasima}},\ }\bibfield  {title} {\enquote {\bibinfo {title} {Molecular
  theory of surface tension},}\ }\href@noop {} {\bibfield  {journal} {\bibinfo
  {journal} {Advances in Chemical Physics}\ }\textbf {\bibinfo {volume} {1}},\
  \bibinfo {pages} {203--237} (\bibinfo {year} {1958})}\BibitemShut {NoStop}%
\bibitem [{\citenamefont {Schofield}\ and\ \citenamefont
  {Henderson}(1982)}]{Schofield1982aa}%
  \BibitemOpen
  \bibfield  {author} {\bibinfo {author} {\bibfnamefont {P.}~\bibnamefont
  {Schofield}}\ and\ \bibinfo {author} {\bibfnamefont {J.~R.}\ \bibnamefont
  {Henderson}},\ }\bibfield  {title} {\enquote {\bibinfo {title} {{Statistical
  Mechanics of Inhomogeneous Fluids}},}\ }\href {\doibase
  10.1098/rspa.1982.0015} {\bibfield  {journal} {\bibinfo  {journal}
  {Proceedings of the Royal Society A: Mathematical, Physical and Engineering
  Sciences}\ }\textbf {\bibinfo {volume} {379}},\ \bibinfo {pages} {231--246}
  (\bibinfo {year} {1982})}\BibitemShut {NoStop}%
\bibitem [{\citenamefont {Todd}\ and\ \citenamefont
  {Daivis}(2007)}]{Todd2007aa}%
  \BibitemOpen
  \bibfield  {author} {\bibinfo {author} {\bibfnamefont {B.~D.}\ \bibnamefont
  {Todd}}\ and\ \bibinfo {author} {\bibfnamefont {P.~J.}\ \bibnamefont
  {Daivis}},\ }\href {\doibase 10.1080/08927020601026629} {\emph {\bibinfo
  {title} {Molecular Simulation}}},\ Vol.~\bibinfo {volume} {33}\ (\bibinfo
  {year} {2007})\ pp.\ \bibinfo {pages} {189--229}\BibitemShut {NoStop}%
\bibitem [{\citenamefont {Evans}\ and\ \citenamefont
  {Morriss}(2008)}]{Evans2008aa}%
  \BibitemOpen
  \bibfield  {author} {\bibinfo {author} {\bibfnamefont {D.~J.}\ \bibnamefont
  {Evans}}\ and\ \bibinfo {author} {\bibfnamefont {G.}~\bibnamefont
  {Morriss}},\ }\href {http://books.google.co.uk/books?id=65URS{\_}vPwuQC}
  {\emph {\bibinfo {title} {{Statistical Mechanics of Nonequilibrium
  Liquids}}}},\ Theoretical chemistry\ (\bibinfo  {publisher} {Cambridge
  University Press},\ \bibinfo {year} {2008})\BibitemShut {NoStop}%
\bibitem [{\citenamefont {Derks}\ \emph {et~al.}(2006)\citenamefont {Derks},
  \citenamefont {Aarts}, \citenamefont {Bonn}, \citenamefont {Lekkerkerker},\
  and\ \citenamefont {Imhof}}]{Derks2006aa}%
  \BibitemOpen
  \bibfield  {author} {\bibinfo {author} {\bibfnamefont {D.}~\bibnamefont
  {Derks}}, \bibinfo {author} {\bibfnamefont {D.~G. A.~L.}\ \bibnamefont
  {Aarts}}, \bibinfo {author} {\bibfnamefont {D.}~\bibnamefont {Bonn}},
  \bibinfo {author} {\bibfnamefont {H.~N.~W.}\ \bibnamefont {Lekkerkerker}}, \
  and\ \bibinfo {author} {\bibfnamefont {A.}~\bibnamefont {Imhof}},\ }\bibfield
   {title} {\enquote {\bibinfo {title} {{Suppression of thermally excited
  capillary waves by shear flow}},}\ }\href {\doibase
  10.1103/PhysRevLett.97.038301} {\bibfield  {journal} {\bibinfo  {journal}
  {Physical Review Letters}\ }\textbf {\bibinfo {volume} {97}} (\bibinfo {year}
  {2006}),\ 10.1103/PhysRevLett.97.038301}\BibitemShut {NoStop}%
\bibitem [{\citenamefont {Thi{\'{e}}baud}\ and\ \citenamefont
  {Bickel}(2010)}]{Thiebaud2010aa}%
  \BibitemOpen
  \bibfield  {author} {\bibinfo {author} {\bibfnamefont {M.}~\bibnamefont
  {Thi{\'{e}}baud}}\ and\ \bibinfo {author} {\bibfnamefont {T.}~\bibnamefont
  {Bickel}},\ }\bibfield  {title} {\enquote {\bibinfo {title} {{Nonequilibrium
  fluctuations of an interface under shear}},}\ }\href {\doibase
  10.1103/PhysRevE.81.031602} {\bibfield  {journal} {\bibinfo  {journal}
  {Physical Review E}\ }\textbf {\bibinfo {volume} {81}},\ \bibinfo {pages}
  {1--13} (\bibinfo {year} {2010})}\BibitemShut {NoStop}%
\bibitem [{\citenamefont {Francois}\ \emph {et~al.}(2017)\citenamefont
  {Francois}, \citenamefont {Xia}, \citenamefont {Punzmann}, \citenamefont
  {Fontana},\ and\ \citenamefont {Shats}}]{Francois2017aa}%
  \BibitemOpen
  \bibfield  {author} {\bibinfo {author} {\bibfnamefont {N.}~\bibnamefont
  {Francois}}, \bibinfo {author} {\bibfnamefont {H.}~\bibnamefont {Xia}},
  \bibinfo {author} {\bibfnamefont {H.}~\bibnamefont {Punzmann}}, \bibinfo
  {author} {\bibfnamefont {P.~W.}\ \bibnamefont {Fontana}}, \ and\ \bibinfo
  {author} {\bibfnamefont {M.}~\bibnamefont {Shats}},\ }\bibfield  {title}
  {\enquote {\bibinfo {title} {{Wave-based liquid-interface metamaterials}},}\
  }\href {\doibase 10.1038/ncomms14325} {\bibfield  {journal} {\bibinfo
  {journal} {Nature Communications}\ }\textbf {\bibinfo {volume} {8}},\
  \bibinfo {pages} {14325} (\bibinfo {year} {2017})}\BibitemShut {NoStop}%
\bibitem [{\citenamefont {Miniewicz}\ \emph {et~al.}(2016)\citenamefont
  {Miniewicz}, \citenamefont {Bartkiewicz}, \citenamefont {Orlikowska},\ and\
  \citenamefont {Dradrach}}]{Miniewicz2016aa}%
  \BibitemOpen
  \bibfield  {author} {\bibinfo {author} {\bibfnamefont {A.}~\bibnamefont
  {Miniewicz}}, \bibinfo {author} {\bibfnamefont {S.}~\bibnamefont
  {Bartkiewicz}}, \bibinfo {author} {\bibfnamefont {H.}~\bibnamefont
  {Orlikowska}}, \ and\ \bibinfo {author} {\bibfnamefont {K.}~\bibnamefont
  {Dradrach}},\ }\bibfield  {title} {\enquote {\bibinfo {title} {{Marangoni
  effect visualized in two-dimensions Optical tweezers for gas bubbles}},}\
  }\href {\doibase 10.1038/srep34787} {\bibfield  {journal} {\bibinfo
  {journal} {Scientific Reports}\ }\textbf {\bibinfo {volume} {6}},\ \bibinfo
  {pages} {34787} (\bibinfo {year} {2016})}\BibitemShut {NoStop}%
\bibitem [{\citenamefont {Smith}\ \emph {et~al.}(2012)\citenamefont {Smith},
  \citenamefont {Heyes}, \citenamefont {Dini},\ and\ \citenamefont
  {Zaki}}]{Smith2012aa}%
  \BibitemOpen
  \bibfield  {author} {\bibinfo {author} {\bibfnamefont {E.~R.}\ \bibnamefont
  {Smith}}, \bibinfo {author} {\bibfnamefont {D.~M.}\ \bibnamefont {Heyes}},
  \bibinfo {author} {\bibfnamefont {D.}~\bibnamefont {Dini}}, \ and\ \bibinfo
  {author} {\bibfnamefont {T.~a.}\ \bibnamefont {Zaki}},\ }\bibfield  {title}
  {\enquote {\bibinfo {title} {{Control-volume representation of molecular
  dynamics}},}\ }\href {\doibase 10.1103/PhysRevE.85.056705} {\bibfield
  {journal} {\bibinfo  {journal} {Physical Review E}\ }\textbf {\bibinfo
  {volume} {85}} (\bibinfo {year} {2012}),\
  10.1103/PhysRevE.85.056705}\BibitemShut {NoStop}%
\bibitem [{\citenamefont {Hardy}(1982)}]{Hardy1982aa}%
  \BibitemOpen
  \bibfield  {author} {\bibinfo {author} {\bibfnamefont {R.~J.}\ \bibnamefont
  {Hardy}},\ }\bibfield  {title} {\enquote {\bibinfo {title} {{Formulas for
  determining local properties in molecular-dynamics simulations : Shock
  waves}},}\ }\href@noop {} {\bibfield  {journal} {\bibinfo  {journal} {Journal
  of Chemical Physics}\ }\textbf {\bibinfo {volume} {76}},\ \bibinfo {pages}
  {622} (\bibinfo {year} {1982})}\BibitemShut {NoStop}%
\bibitem [{\citenamefont {Lutsko}(1988)}]{Lutsko1988aa}%
  \BibitemOpen
  \bibfield  {author} {\bibinfo {author} {\bibfnamefont {J.~F.}\ \bibnamefont
  {Lutsko}},\ }\bibfield  {title} {\enquote {\bibinfo {title} {{Stress and
  elastic constants in anisotropic solids: Molecular dynamics techniques}},}\
  }\href {\doibase 10.1063/1.341877} {\bibfield  {journal} {\bibinfo  {journal}
  {Journal of Applied Physics}\ }\textbf {\bibinfo {volume} {64}},\ \bibinfo
  {pages} {1152--1154} (\bibinfo {year} {1988})},\ \Eprint
  {http://arxiv.org/abs/arXiv:1011.1669v3} {arXiv:1011.1669v3} \BibitemShut
  {NoStop}%
\bibitem [{\citenamefont {Trokhymchuk}\ and\ \citenamefont
  {Alejandre}(1999)}]{Trokhymchuk1999aa}%
  \BibitemOpen
  \bibfield  {author} {\bibinfo {author} {\bibfnamefont {A.}~\bibnamefont
  {Trokhymchuk}}\ and\ \bibinfo {author} {\bibfnamefont {J.}~\bibnamefont
  {Alejandre}},\ }\bibfield  {title} {\enquote {\bibinfo {title} {{Computer
  simulations of liquid/vapor interface in Lennard-Jones fluids: Some questions
  and answers}},}\ }\href {\doibase 10.1063/1.480192} {\bibfield  {journal}
  {\bibinfo  {journal} {The Journal of Chemical Physics}\ }\textbf {\bibinfo
  {volume} {111}},\ \bibinfo {pages} {8510} (\bibinfo {year}
  {1999})}\BibitemShut {NoStop}%
\bibitem [{\citenamefont {Bresme}\ \emph {et~al.}(2008)\citenamefont {Bresme},
  \citenamefont {Chac{\'{o}}n}, \citenamefont {Tarazona},\ and\ \citenamefont
  {Tay}}]{Bresme2008aa}%
  \BibitemOpen
  \bibfield  {author} {\bibinfo {author} {\bibfnamefont {F.}~\bibnamefont
  {Bresme}}, \bibinfo {author} {\bibfnamefont {E.}~\bibnamefont
  {Chac{\'{o}}n}}, \bibinfo {author} {\bibfnamefont {P.}~\bibnamefont
  {Tarazona}}, \ and\ \bibinfo {author} {\bibfnamefont {K.}~\bibnamefont
  {Tay}},\ }\bibfield  {title} {\enquote {\bibinfo {title} {{Intrinsic
  Structure of Hydrophobic Surfaces: The Oil-Water Interface}},}\ }\href
  {\doibase 10.1103/PhysRevLett.101.056102} {\bibfield  {journal} {\bibinfo
  {journal} {Physical Review Letters}\ }\textbf {\bibinfo {volume} {101}},\
  \bibinfo {pages} {056102} (\bibinfo {year} {2008})}\BibitemShut {NoStop}%
\bibitem [{\citenamefont {Mastny}\ and\ \citenamefont
  {de~Pablo}(2007)}]{Mastny2007aa}%
  \BibitemOpen
  \bibfield  {author} {\bibinfo {author} {\bibfnamefont {E.~A.}\ \bibnamefont
  {Mastny}}\ and\ \bibinfo {author} {\bibfnamefont {J.~J.}\ \bibnamefont
  {de~Pablo}},\ }\bibfield  {title} {\enquote {\bibinfo {title} {Melting line
  of the lennard-jones system, infinite size, and full potential},}\ }\href
  {\doibase http://dx.doi.org/10.1063/1.2753149} {\bibfield  {journal}
  {\bibinfo  {journal} {The Journal of Chemical Physics}\ }\textbf {\bibinfo
  {volume} {127}},\ \bibinfo {eid} {104504} (\bibinfo {year} {2007}),\
  http://dx.doi.org/10.1063/1.2753149}\BibitemShut {NoStop}%
\bibitem [{\citenamefont {Errington}, \citenamefont {Debenedetti},\ and\
  \citenamefont {Torquato}(2003)}]{Errington2003aa}%
  \BibitemOpen
  \bibfield  {author} {\bibinfo {author} {\bibfnamefont {J.~R.}\ \bibnamefont
  {Errington}}, \bibinfo {author} {\bibfnamefont {P.~G.}\ \bibnamefont
  {Debenedetti}}, \ and\ \bibinfo {author} {\bibfnamefont {S.}~\bibnamefont
  {Torquato}},\ }\bibfield  {title} {\enquote {\bibinfo {title} {Quantification
  of order in the lennard-jones system},}\ }\href {\doibase
  http://dx.doi.org/10.1063/1.1532344} {\bibfield  {journal} {\bibinfo
  {journal} {The Journal of Chemical Physics}\ }\textbf {\bibinfo {volume}
  {118}},\ \bibinfo {pages} {2256--2263} (\bibinfo {year} {2003})}\BibitemShut
  {NoStop}%
\bibitem [{\citenamefont {Braga}\ \emph {et~al.}(2016)\citenamefont {Braga},
  \citenamefont {Muscatello}, \citenamefont {Lau}, \citenamefont {M{\"u}ller},\
  and\ \citenamefont {Jackson}}]{Braga2016aa}%
  \BibitemOpen
  \bibfield  {author} {\bibinfo {author} {\bibfnamefont {C.}~\bibnamefont
  {Braga}}, \bibinfo {author} {\bibfnamefont {J.}~\bibnamefont {Muscatello}},
  \bibinfo {author} {\bibfnamefont {G.}~\bibnamefont {Lau}}, \bibinfo {author}
  {\bibfnamefont {E.~A.}\ \bibnamefont {M{\"u}ller}}, \ and\ \bibinfo {author}
  {\bibfnamefont {G.}~\bibnamefont {Jackson}},\ }\bibfield  {title} {\enquote
  {\bibinfo {title} {Nonequilibrium study of the intrinsic free-energy profile
  across a liquid-vapour interface},}\ }\href {\doibase 10.1063/1.4940137}
  {\bibfield  {journal} {\bibinfo  {journal} {The Journal of Chemical Physics}\
  }\textbf {\bibinfo {volume} {144}},\ \bibinfo {pages} {044703} (\bibinfo
  {year} {2016})}\BibitemShut {NoStop}%
\bibitem [{\citenamefont {Plimpton}(1995)}]{Plimpton1995aa}%
  \BibitemOpen
  \bibfield  {author} {\bibinfo {author} {\bibfnamefont {S.}~\bibnamefont
  {Plimpton}},\ }\bibfield  {title} {\enquote {\bibinfo {title} {{Fast Parallel
  Algorithms for Short Range Molecular Dynamics}},}\ }\href
  {http://citeseerx.ist.psu.edu/viewdoc/download?doi=10.1.1.35.6047\&rep=rep1\&type=pdf}
  {\bibfield  {journal} {\bibinfo  {journal} {Journal of Computational
  Physics}\ }\textbf {\bibinfo {volume} {117}},\ \bibinfo {pages} {1--19}
  (\bibinfo {year} {1995})}\BibitemShut {NoStop}%
\bibitem [{\citenamefont {Hoover}(1985)}]{Hoover1985aa}%
  \BibitemOpen
  \bibfield  {author} {\bibinfo {author} {\bibfnamefont {W.}~\bibnamefont
  {Hoover}},\ }\bibfield  {title} {\enquote {\bibinfo {title} {{Canonical
  dynamics: equilibrium phase-space distributions}},}\ }\href
  {http://pra.aps.org/abstract/PRA/v31/i3/p1695\_1} {\bibfield  {journal}
  {\bibinfo  {journal} {Physical Review A}\ }\textbf {\bibinfo {volume} {31}},\
  \bibinfo {pages} {1695--1697} (\bibinfo {year} {1985})}\BibitemShut {NoStop}%
\bibitem [{\citenamefont {Nos{\'e}}(1984)}]{Nose1984aa}%
  \BibitemOpen
  \bibfield  {author} {\bibinfo {author} {\bibfnamefont {S.}~\bibnamefont
  {Nos{\'e}}},\ }\bibfield  {title} {\enquote {\bibinfo {title} {A molecular
  dynamics method for simulations in the canonical ensemble},}\ }\href
  {\doibase 10.1080/00268978400101201} {\bibfield  {journal} {\bibinfo
  {journal} {Molecular Physics}\ }\textbf {\bibinfo {volume} {52}},\ \bibinfo
  {pages} {255--268} (\bibinfo {year} {1984})}\BibitemShut {NoStop}%
\bibitem [{\citenamefont {Stillinger}(1982)}]{Stillinger1982ab}%
  \BibitemOpen
  \bibfield  {author} {\bibinfo {author} {\bibfnamefont {F.~H.}\ \bibnamefont
  {Stillinger}},\ }\bibfield  {title} {\enquote {\bibinfo {title} {Capillary
  waves and the inherent density profile for the liquid--vapor interface},}\
  }\href {\doibase 10.1063/1.443075} {\bibfield  {journal} {\bibinfo  {journal}
  {The Journal of Chemical Physics}\ }\textbf {\bibinfo {volume} {76}},\
  \bibinfo {pages} {1087--1091} (\bibinfo {year} {1982})}\BibitemShut {NoStop}%
\bibitem [{\citenamefont {Chac{\'{o}}n}\ and\ \citenamefont
  {Tarazona}(2003)}]{Chacon2003aa}%
  \BibitemOpen
  \bibfield  {author} {\bibinfo {author} {\bibfnamefont {E.}~\bibnamefont
  {Chac{\'{o}}n}}\ and\ \bibinfo {author} {\bibfnamefont {P.}~\bibnamefont
  {Tarazona}},\ }\bibfield  {title} {\enquote {\bibinfo {title} {{Intrinsic
  Profiles beyond the Capillary Wave Theory: A Monte Carlo Study}},}\ }\href
  {\doibase 10.1103/PhysRevLett.91.166103} {\bibfield  {journal} {\bibinfo
  {journal} {Physical Review Letters}\ }\textbf {\bibinfo {volume} {91}},\
  \bibinfo {pages} {166103} (\bibinfo {year} {2003})}\BibitemShut {NoStop}%
\bibitem [{\citenamefont {Tarazona}\ and\ \citenamefont
  {Chac{\'{o}}n}(2004)}]{Tarazona2004aa}%
  \BibitemOpen
  \bibfield  {author} {\bibinfo {author} {\bibfnamefont {P.}~\bibnamefont
  {Tarazona}}\ and\ \bibinfo {author} {\bibfnamefont {E.}~\bibnamefont
  {Chac{\'{o}}n}},\ }\bibfield  {title} {\enquote {\bibinfo {title} {{Monte
  Carlo intrinsic surfaces and density profiles for liquid surfaces}},}\ }\href
  {\doibase 10.1103/PhysRevB.70.235407} {\bibfield  {journal} {\bibinfo
  {journal} {Physical Review B}\ }\textbf {\bibinfo {volume} {70}},\ \bibinfo
  {pages} {235407} (\bibinfo {year} {2004})}\BibitemShut {NoStop}%
\bibitem [{\citenamefont {Chowdhary}\ and\ \citenamefont
  {Ladanyi}(2008)}]{Chowdhary2008aa}%
  \BibitemOpen
  \bibfield  {author} {\bibinfo {author} {\bibfnamefont {J.}~\bibnamefont
  {Chowdhary}}\ and\ \bibinfo {author} {\bibfnamefont {B.~M.}\ \bibnamefont
  {Ladanyi}},\ }\bibfield  {title} {\enquote {\bibinfo {title} {{Surface
  fluctuations at the liquid-liquid interface}},}\ }\href {\doibase
  10.1103/PhysRevE.77.031609} {\bibfield  {journal} {\bibinfo  {journal}
  {Physical Review E}\ }\textbf {\bibinfo {volume} {77}},\ \bibinfo {pages}
  {1--14} (\bibinfo {year} {2008})}\BibitemShut {NoStop}%
\bibitem [{\citenamefont {Jorge}\ and\ \citenamefont
  {Cordeiro}(2007)}]{Jorge2007aa}%
  \BibitemOpen
  \bibfield  {author} {\bibinfo {author} {\bibfnamefont {M.}~\bibnamefont
  {Jorge}}\ and\ \bibinfo {author} {\bibfnamefont {M.~N. D.~S.}\ \bibnamefont
  {Cordeiro}},\ }\bibfield  {title} {\enquote {\bibinfo {title} {Intrinsic
  structure and dynamics of the water/nitrobenzene interface},}\ }\href
  {\doibase 10.1021/jp076178q} {\bibfield  {journal} {\bibinfo  {journal} {The
  Journal of Physical Chemistry C}\ }\textbf {\bibinfo {volume} {111}},\
  \bibinfo {pages} {17612--17626} (\bibinfo {year} {2007})}\BibitemShut
  {NoStop}%
\bibitem [{\citenamefont {P{\'a}rtay}\ \emph {et~al.}(2008)\citenamefont
  {P{\'a}rtay}, \citenamefont {Hantal}, \citenamefont {Jedlovszky},
  \citenamefont {Vincze},\ and\ \citenamefont {Horvai}}]{Partay2008ab}%
  \BibitemOpen
  \bibfield  {author} {\bibinfo {author} {\bibfnamefont {L.~B.}\ \bibnamefont
  {P{\'a}rtay}}, \bibinfo {author} {\bibfnamefont {G.}~\bibnamefont {Hantal}},
  \bibinfo {author} {\bibfnamefont {P.}~\bibnamefont {Jedlovszky}}, \bibinfo
  {author} {\bibfnamefont {{\'A}.}~\bibnamefont {Vincze}}, \ and\ \bibinfo
  {author} {\bibfnamefont {G.}~\bibnamefont {Horvai}},\ }\bibfield  {title}
  {\enquote {\bibinfo {title} {A new method for determining the interfacial
  molecules and characterizing the surface roughness in computer simulations.
  application to the liquid--vapor interface of water},}\ }\href {\doibase
  10.1002/jcc.20852} {\bibfield  {journal} {\bibinfo  {journal} {Journal of
  Computational Chemistry}\ }\textbf {\bibinfo {volume} {29}},\ \bibinfo
  {pages} {945--956} (\bibinfo {year} {2008})}\BibitemShut {NoStop}%
\bibitem [{\citenamefont {Jorge}, \citenamefont {Jedlovszky},\ and\
  \citenamefont {Cordeiro}(2010)}]{Jorge2010ab}%
  \BibitemOpen
  \bibfield  {author} {\bibinfo {author} {\bibfnamefont {M.}~\bibnamefont
  {Jorge}}, \bibinfo {author} {\bibfnamefont {P.}~\bibnamefont {Jedlovszky}}, \
  and\ \bibinfo {author} {\bibfnamefont {M.~N. D.~S.}\ \bibnamefont
  {Cordeiro}},\ }\bibfield  {title} {\enquote {\bibinfo {title} {A critical
  assessment of methods for the intrinsic analysis of liquid interfaces. 1.
  surface site distributions},}\ }\href {\doibase 10.1021/jp101035r} {\bibfield
   {journal} {\bibinfo  {journal} {The Journal of Physical Chemistry C}\
  }\textbf {\bibinfo {volume} {114}},\ \bibinfo {pages} {11169--11179}
  (\bibinfo {year} {2010})}\BibitemShut {NoStop}%
\bibitem [{\citenamefont {Sega}\ \emph {et~al.}(2013)\citenamefont {Sega},
  \citenamefont {Kantorovich}, \citenamefont {Jedlovszky},\ and\ \citenamefont
  {Jorge}}]{Sega2013aa}%
  \BibitemOpen
  \bibfield  {author} {\bibinfo {author} {\bibfnamefont {M.}~\bibnamefont
  {Sega}}, \bibinfo {author} {\bibfnamefont {S.~S.}\ \bibnamefont
  {Kantorovich}}, \bibinfo {author} {\bibfnamefont {P.}~\bibnamefont
  {Jedlovszky}}, \ and\ \bibinfo {author} {\bibfnamefont {M.}~\bibnamefont
  {Jorge}},\ }\bibfield  {title} {\enquote {\bibinfo {title} {The generalized
  identification of truly interfacial molecules (itim) algorithm for nonplanar
  interfaces},}\ }\href {\doibase 10.1063/1.4776196} {\bibfield  {journal}
  {\bibinfo  {journal} {The Journal of Chemical Physics}\ }\textbf {\bibinfo
  {volume} {138}},\ \bibinfo {pages} {044110} (\bibinfo {year}
  {2013})}\BibitemShut {NoStop}%
\bibitem [{\citenamefont {Willard}\ and\ \citenamefont
  {Chandler}(2010)}]{Willard2010aa}%
  \BibitemOpen
  \bibfield  {author} {\bibinfo {author} {\bibfnamefont {A.~P.}\ \bibnamefont
  {Willard}}\ and\ \bibinfo {author} {\bibfnamefont {D.}~\bibnamefont
  {Chandler}},\ }\bibfield  {title} {\enquote {\bibinfo {title} {Instantaneous
  liquid interfaces},}\ }\href {\doibase 10.1021/jp909219k} {\bibfield
  {journal} {\bibinfo  {journal} {The Journal of Physical Chemistry B}\
  }\textbf {\bibinfo {volume} {114}},\ \bibinfo {pages} {1954--1958} (\bibinfo
  {year} {2010})},\ \bibinfo {note} {pMID: 20055377}\BibitemShut {NoStop}%
\bibitem [{\citenamefont {Rousseeuw}\ and\ \citenamefont
  {Leroy}(2005)}]{Rousseeuw2005aa}%
  \BibitemOpen
  \bibfield  {author} {\bibinfo {author} {\bibfnamefont {P.}~\bibnamefont
  {Rousseeuw}}\ and\ \bibinfo {author} {\bibfnamefont {A.}~\bibnamefont
  {Leroy}},\ }\href@noop {} {\emph {\bibinfo {title} {Robust Regression and
  Outlier Detection}}},\ Wiley Series in Probability and Statistics\ (\bibinfo
  {publisher} {Wiley},\ \bibinfo {year} {2005})\BibitemShut {NoStop}%
\bibitem [{\citenamefont {Chac{\'{o}}n}\ \emph {et~al.}(2009)\citenamefont
  {Chac{\'{o}}n}, \citenamefont {Fern{\'{a}}ndez}, \citenamefont {Duque},
  \citenamefont {Delgado-Buscalioni},\ and\ \citenamefont
  {Tarazona}}]{Chacon2009aa}%
  \BibitemOpen
  \bibfield  {author} {\bibinfo {author} {\bibfnamefont {E.}~\bibnamefont
  {Chac{\'{o}}n}}, \bibinfo {author} {\bibfnamefont {E.}~\bibnamefont
  {Fern{\'{a}}ndez}}, \bibinfo {author} {\bibfnamefont {D.}~\bibnamefont
  {Duque}}, \bibinfo {author} {\bibfnamefont {R.}~\bibnamefont
  {Delgado-Buscalioni}}, \ and\ \bibinfo {author} {\bibfnamefont
  {P.}~\bibnamefont {Tarazona}},\ }\bibfield  {title} {\enquote {\bibinfo
  {title} {{Comparative study of the surface layer density of liquid
  surfaces}},}\ }\href {\doibase 10.1103/PhysRevB.80.195403} {\bibfield
  {journal} {\bibinfo  {journal} {Physical Review B}\ }\textbf {\bibinfo
  {volume} {80}},\ \bibinfo {pages} {195403} (\bibinfo {year}
  {2009})}\BibitemShut {NoStop}%
\bibitem [{\citenamefont {Chac{\'{o}}n}\ and\ \citenamefont
  {Tarazona}(2005)}]{Chacon2005aa}%
  \BibitemOpen
  \bibfield  {author} {\bibinfo {author} {\bibfnamefont {E.}~\bibnamefont
  {Chac{\'{o}}n}}\ and\ \bibinfo {author} {\bibfnamefont {P.}~\bibnamefont
  {Tarazona}},\ }\bibfield  {title} {\enquote {\bibinfo {title}
  {{Characterization of the intrinsic density profiles for liquid surfaces}},}\
  }\href {\doibase 10.1088/0953-8984/17/45/039} {\bibfield  {journal} {\bibinfo
   {journal} {Journal of Physics: Condensed Matter}\ }\textbf {\bibinfo
  {volume} {17}},\ \bibinfo {pages} {S3493--S3498} (\bibinfo {year}
  {2005})}\BibitemShut {NoStop}%
\bibitem [{\citenamefont {Kirkwood}\ and\ \citenamefont
  {Buff}(1949)}]{Kirkwood1949aa}%
  \BibitemOpen
  \bibfield  {author} {\bibinfo {author} {\bibfnamefont {J.~G.}\ \bibnamefont
  {Kirkwood}}\ and\ \bibinfo {author} {\bibfnamefont {F.~P.}\ \bibnamefont
  {Buff}},\ }\bibfield  {title} {\enquote {\bibinfo {title} {{The Statistical
  Mechanical Theory of Surface Tension}},}\ }\href {\doibase 10.1063/1.1747248}
  {\bibfield  {journal} {\bibinfo  {journal} {The Journal of Chemical Physics}\
  }\textbf {\bibinfo {volume} {17}},\ \bibinfo {pages} {338} (\bibinfo {year}
  {1949})}\BibitemShut {NoStop}%
\bibitem [{\citenamefont {Kirkwood}\ and\ \citenamefont
  {Buff}(1951)}]{Kirkwood1951aa}%
  \BibitemOpen
  \bibfield  {author} {\bibinfo {author} {\bibfnamefont {J.~G.}\ \bibnamefont
  {Kirkwood}}\ and\ \bibinfo {author} {\bibfnamefont {F.~P.}\ \bibnamefont
  {Buff}},\ }\bibfield  {title} {\enquote {\bibinfo {title} {{The Statistical
  Mechanical Theory of Solutions. I}},}\ }\href {\doibase 10.1063/1.1748352}
  {\bibfield  {journal} {\bibinfo  {journal} {The Journal of Chemical Physics}\
  }\textbf {\bibinfo {volume} {19}},\ \bibinfo {pages} {774} (\bibinfo {year}
  {1951})}\BibitemShut {NoStop}%
\bibitem [{\citenamefont {Yatsyshin}\ \emph {et~al.}(2017)\citenamefont
  {Yatsyshin}, \citenamefont {Parry}, \citenamefont {Rasc\'on},\ and\
  \citenamefont {Kalliadasis}}]{Peter2017}%
  \BibitemOpen
  \bibfield  {author} {\bibinfo {author} {\bibfnamefont {P.}~\bibnamefont
  {Yatsyshin}}, \bibinfo {author} {\bibfnamefont {A.~O.}~\bibnamefont {Parry}},
  \bibinfo {author} {\bibfnamefont {C.}~\bibnamefont {Rasc\'on}}, \ and\
  \bibinfo {author} {\bibfnamefont {S.}~\bibnamefont {Kalliadasis}},\
  }\bibfield  {title} {\enquote {\bibinfo {title} {Classical density functional
  study of wetting transitions on nanopatterned surfaces},}\ }\href
  {http://iopscience.iop.org/article/10.1088/1361-648X/aa4fd7/meta} {\bibfield
  {journal} {\bibinfo  {journal} {J. Phys.: Condens. Matter}\ }\textbf
  {\bibinfo {volume} {29}},\ \bibinfo {pages} {094001} (\bibinfo {year}
  {2017})}\BibitemShut {NoStop}%
\bibitem [{\citenamefont {Morciano}\ \emph {et~al.}(2017)\citenamefont
  {Morciano}, \citenamefont {Fasano}, \citenamefont {Nold}, \citenamefont
  {Braga}, \citenamefont {Yatsyshin}, \citenamefont {Sibley}, \citenamefont
  {Goddard}, \citenamefont {Chiavazzo}, \citenamefont {Asinari},\ and\
  \citenamefont {Kalliadasis}}]{Matteo2017}%
  \BibitemOpen
  \bibfield  {author} {\bibinfo {author} {\bibfnamefont {M.}~\bibnamefont
  {Morciano}}, \bibinfo {author} {\bibfnamefont {M.}~\bibnamefont {Fasano}},
  \bibinfo {author} {\bibfnamefont {A.}~\bibnamefont {Nold}}, \bibinfo {author}
  {\bibfnamefont {C.}~\bibnamefont {Braga}}, \bibinfo {author} {\bibfnamefont
  {P.}~\bibnamefont {Yatsyshin}}, \bibinfo {author} {\bibfnamefont
  {D.}~\bibnamefont {Sibley}}, \bibinfo {author} {\bibfnamefont
  {B.}~\bibnamefont {Goddard}}, \bibinfo {author} {\bibfnamefont
  {E.}~\bibnamefont {Chiavazzo}}, \bibinfo {author} {\bibfnamefont
  {P.}~\bibnamefont {Asinari}}, \ and\ \bibinfo {author} {\bibfnamefont
  {S.}~\bibnamefont {Kalliadasis}},\ }\bibfield  {title} {\enquote {\bibinfo
  {title} {Nonequilibrium molecular dynamics simulations of nanoconfined fluids
  at solid-liquid interfaces},}\ }\href {https://doi.org/10.1063/1.4986904}
  {\bibfield  {journal} {\bibinfo  {journal} {J. Chem. Phys.}\ }\textbf
  {\bibinfo {volume} {146}},\ \bibinfo {pages} {244507} (\bibinfo {year}
  {2017})}\BibitemShut {NoStop}%
\end{thebibliography}

%
%\begin{thebibliography}{10}
%\bibitem{Kirkwood1935}
%J.~G. Kirkwood,
%\newblock The Journal of Chemical Physics {\bf 3}, 300 (1935).
%\end{thebibliography}
\end{document}